\documentclass[varenna]{cimento}
\usepackage{graphicx}
\usepackage{amsmath}
\usepackage{amssymb}

\newcommand{\RNS}{$R_{\rm NS}$}
\newcommand{\be}{\begin{equation}}
\newcommand{\ee}{\end{equation}}
\title{Particle Acceleration in Pulsars and Pulsar Wind Nebulae}
\author{Elena Amato}
\institute{INAF - Osservatorio Astrofisico di Arcetri, Largo E. Fermi 5, 50125, Firenze, Italy}

\shortauthor{E.~Amato}

\begin{document}
\maketitle


\begin{abstract}
These notes summarise the contents of the lectures I delivered at the International School of Physics ``Enrico Fermi'' on ``Foundations of Cosmic Ray Astrophysics''. The lectures were dealing with the physics of Pulsars and Pulsar Wind Nebulae (PWNe) in the Cosmic Ray (CR) perspective. 

It has become now clear that the processes taking place in the environment of fast rotating, highly magnetized neutron stars, often detected as pulsars, play a crucial role in the formation of the CR spectrum detected at the Earth. These lectures discuss the main aspects of this connection. Pulsars are likely contributors of the CR lepton flux at the Earth thanks to their nature of electron-positron factories. Pulsars and their nebulae are the best potential leptonic PeVatron in the Galaxy, and the Crab Nebula, the prototype of the Pulsar Wind Nebula class is the only established PeVatron in the Galaxy. Pulsars are however also potential sources of high energy hadrons, up to the energies relevant for UHECRs. Pulsars and their nebulae are the best potential leptonic PeVatrons in the Galaxy, and the Crab Nebula, the prototype of the Pulsar Wind Nebula class, is the only established PeVatron in the Galaxy. Finally regions of suppressed particle diffusion have been observed around evolved pulsars, the so-called {\it TeV halos}, which could have an impact on galactic CR transport.

These lectures discuss the physics of pulsars and PWNe, summarising what we know about these systems and what pieces of information are still missing to fully assess their role in all the above mentioned Cosmic Ray connected aspects.

\end{abstract}

\section{Lecture I: pulsars and their magnetospheres}
\subsection{Brief historical notes}
The existence of neutron stars (NS hereafter) was first proposed in 1934, only one year after the discovery of the neutron, and at the same time as the association between supernovae and Cosmic Rays (CRs in the following). At that time, in two articles rich of incredibly pioneering suggestions, Baade and Zwicky \cite{BaadeZwicky1,BaadeZwicky2} proposed that supernova explosions were powered by the neutronization of matter in the transition of an ordinary star to a neutron star, and that the same process would cause CR acceleration. 

It was clear from the very beginning that NSs would be tiny objects, extremely difficult to observe. The first theoretical estimate of the mass of a NS, put forward in 1939 by Tolmann, Oppenheimer and Volkoff \cite{TOV39}, resulted in $M_{\rm NS}=0.7M_\odot$. The subject was long forgotten. A revised mass estimate with values of $M_{\rm NS}$ up to $2M_\odot$ was published in 1959 \cite{Cameron59}, together with an estimate of the radius \RNS$<$10 km. The prospects for direct observation of such an object remained very poor. 

However, in 1967, Pacini \cite{Pacini67} suggested that NSs could make themselves more easily detectable in an indirect way, as energy sources for the non-thermal activity of some classes of Supernova Remnants. More specifically he suggested such a star as the source powering the activity of the Crab Nebula, a bright nebula earlier identified as the remnant of a supernova explosion observed by Chinese astronomers in 1054 A.D. 

The suggestion by Pacini built on the expectation that during the gravitational collapse leading to the formation of a NS both angular momentum and magnetic flux must be conserved: the result would be an increase of both spin frequency and magnetic field by a factor $\left(R_i/R_{\rm NS}\right)^2\approx 10^{10}$, where $R_i\approx 10^6$ km is the radius of the parent ordinary star. The NS would then host a magnetic field $B_{\rm NS}\approx 10^{12}$ G and spin with a period in the millisecond range. 

It is worth noticing, in passing, that such an object is clearly extremely interesting for both quantum physics and general relativity. The stellar magnetic field is close to the QED limit $B_{\rm NS}\approx (0.02-1) B_{\rm QED}$ (where $B_{\rm QED}=m_e^2 c^3/\hbar e=5 \times 10^{13}$ G, is the magnetic field such that the frequency of electron Larmor gyration, $\omega_{ce}=e B/m_e c$, corresponds to a photon energy $\hbar \omega_{ce}$ equal to the electron rest-mass energy, $m_e c^2$). At the same time, the star Schwarzschild radius $R_{GR}=2GM_{\rm NS}/c^2$ (with G the gravitational constant) is a sizeable fraction of the stellar radius, $R_{GR}\approx R_{\rm NS}/3$ (notice that this ratio is $\ll1$ for ordinary matter and reaching 1 only for black holes).
\begin{figure}[h!!!]
\begin{center}
\includegraphics[width=0.25\textwidth]{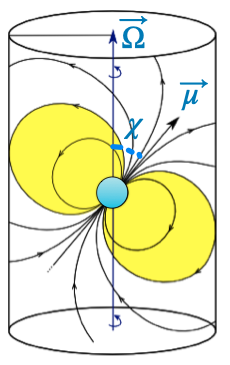}  
\hspace{1cm}
\includegraphics[width=0.65\textwidth]{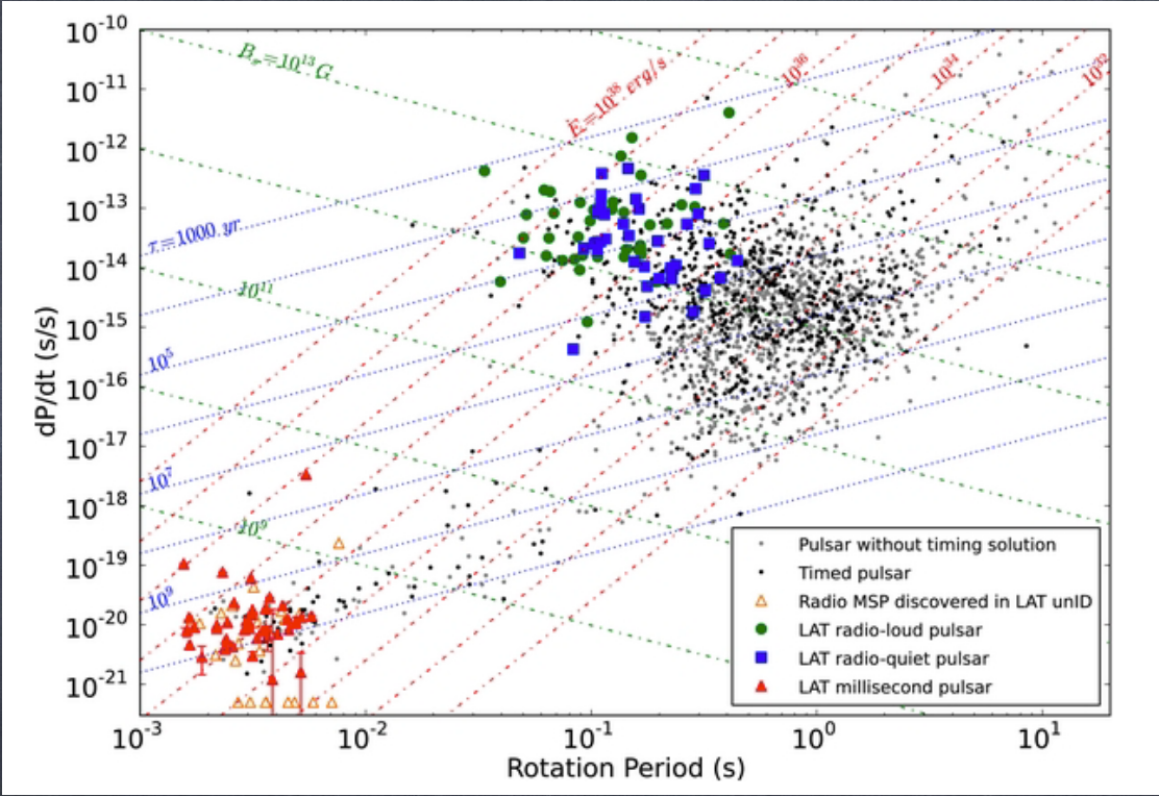}
\caption{\footnotesize{Left panel: sketch of an oblique rotating magnetic dipole. Right panel: $P-\dot P$ diagram for pulsars, from \cite{Abdo13}}}
\label{fig:psrppdot}
\end{center}
\end{figure}


\subsection{The inclined dipole model}
\label{sec:oblique}
The huge rotational energy stored in a fast-spinning NS could easily be released in the form of radiation thanks to the presence of a strong magnetic field. This is the calculation that \cite{Pacini67} carried out.

Let us consider a Cartesian reference frame with axes $({\bf e}_1,\ {\bf e}_2,\ {\bf e}_3)$, and assume that the star spins with frequency $\Omega_\star$ around ${\bf e}_3$, while its magnetic dipole moment $\vec \mu$ makes an angle $\chi$ with ${\bf e}_3$ (see left panel in Fig.~\ref{fig:psrppdot}). The star magnetic field can be written as:
\be
\vec B=\frac{3 {\bf e}_R (\vec \mu \cdot {\bf e}_R )-\vec \mu }{R^3}=B_\star \left(\frac{R_\star}{R}\right)^3 \left[\cos \theta\ {\bf e}_R-\frac{\sin \theta}{2} {\bf e}_\theta\right]
\label{eq:Bstar}
\ee
with
\be
\vec \mu=\mu_\star \left[\sin \chi \left(\cos \Omega t\ {\bf e}_1+\sin \Omega t\ {\bf e}_2 \right)+\cos \chi\ {\bf e_3}\right]\ ,
\label{eq:mudip}
\ee
where $({\bf e}_1,{\bf e}_2, {\bf e}_3)$ and $({\bf e}_R,{\bf e}_\theta, {\bf e}_\phi)$ are orthogonal unit vectors in Cartesian and spherical coordinates respectively.

In the above equations, $t$ is the time, ${\bf e}_R$ is the radial unit vector in spherical coordinates and $\mu_\star=B_\star R_\star^3/2$, with $B_\star$ the field at the star magnetic pole.
The electromagnetic power emitted in this configuration is easily computed by using Larmor formula, $\dot E=(2/3) \ddot \mu^2/c^3$, and reads:
\be
\dot E=\frac{B_\star^2 R_\star^6 \Omega^4 \sin^2 \chi}{6 c^3}=10^{40} \frac{B_{12}^2}{P_{-3}^4} \sin^2 \chi\  {\rm erg/s}
\label{eq:edotdip}
\ee
where $B_{12}$ is the magnetic field at the star pole in units of $10^{12}$ G and $P_{-3}$ is the star rotation period in units of milliseconds. The conclusion was that the power release by such a hypothetical object sitting at the heart of the Crab Nebula would provide a viable explanation for the activity observed in this remnant, including its bright non-thermal emission at radio frequencies \cite{Shklovsky58}. 

Final confirmation of the existence of NSs, however, came only a year later, after the discovery of pulsars in 1968 \cite{HewishBell68}: a pulsating radio source with a period of only $1.33 s$ was serendipitously found by Jocelyn Bell, a PhD student at Cambridge, collecting radio data for her PhD thesis on quasars. The clear identification of the signal origin as extra-terrestrial initially posed a puzzle, because, if linked to rotation, the period was too short to be associated with a white dwarf. Any object of mass $M$ and radius $R$ can only rotate as fast as its break-up frequency, defined by the condition that the centrifugal force equals gravity at the surface: $\Omega ^2 R\leq G M/R^2\, \Rightarrow P\geq P_{min}=2 \pi\sqrt{R^3/GM}$. It is straightforward to see that a period as short as $1s$ implies a size comparable to that of a small planet, and smaller than the typical size of a white dwarf. The possible association with a that far hypothetical neutron star was suggested already in the discovery article, and reinforced by the following detection of a few more sources with similarly short periods \cite{Pilkington68}. The suspicion became a certainty when a PULSAting Radio source was found at the center of the Crab Nebula \cite{Staelin68}, exactly as Pacini had hypothesized. 

\subsection{The $P-\dot P$ diagram}
\label{sec:p-pdot}
Today, the number of known pulsars has grown to several thousands, and these sources are detected not only in radio, but across the entire electromagnetic spectrum, including gamma-rays and, more recently, Very High Energy (VHE hereafter) gamma-rays \cite{CrabMAGICTeV}. It is customary to classify pulsars based on their position on the so-called $P-\dot P$ diagram (right panel of Fig.~\ref{fig:psrppdot}), that is supposed to provide an immediate estimate of the pulsar age (blue lines in the plot), bolometric luminosity (red lines) and magnetic field (green lines). These estimates are based on the assumption that the energy loss is well described by Eq.~\ref{eq:edotdip} and that the ultimate energy reservoir for a pulsar is its rotation, $I_\star \Omega^2/2$ with $I_\star$ the star momentum of inertia. The latter assumption implies:
\be
\dot E= -I_\star \Omega \dot \Omega=4 \pi^2 I_\star \frac{\dot P}{P^3}\ ,
\label{eq:edotrot}
\ee
with $P=2 \pi/\Omega$. 

Equating $\dot E$ from Eq.~\ref{eq:edotdip} and \ref{eq:edotrot}:
\begin{equation}
    \dot \Omega=-\frac{B_\star^2 R_\star^6 \sin^2 \chi}{6 c^3 I_\star}\Omega^3\ ,
    \label{eq:omegadot3}
\end{equation}
which can be expressed in terms of $P$ and $\dot P$ as:
\begin{equation}
P\ \dot P=\frac{(2 \pi)^2 B_\star^2 R_\star^6 \sin^2 \chi}{6 c^3 I_\star\ } .
    \label{eq:ppdot}
\end{equation}
Assuming that the {\it rhs} of Eq.~\ref{eq:ppdot} is a constant (and we shall comment later on the goodness of this assumption), one can use it to derive the pulsar age and the minimum magnetic field (the two quantities represented in the right panel of Fig.~\ref{fig:psrppdot} together with $\dot E$).

The {\bf pulsar characteristic age} is simply obtained integrating Eq.~\ref{eq:ppdot} in time: $P(t)^2-P_0^2=2\ A\ t$ with $A$ defined as the {\it rhs} of the equation. Further assuming that the initial pulsar period is $P_0\ll P(t)$ and recalling that $A=P \dot P$, one obtains
\begin{equation}
t_{\rm ch}=\frac{P}{2 \dot P}\ .
    \label{eq:tchar1}
\end{equation}
Analogously, setting $\sin \chi=1$ in Eq.~\ref{eq:ppdot}, we derive the {\bf minimum magnetic field} of a PSR of given $P$ and $\dot P$ as:
\begin{equation}
    B_{\rm min}=\sqrt{\frac{3 c^3 I_\star}{2 \pi^2}}\ \sqrt{P \dot P}\ .
    \label{eq:bmin}
\end{equation}
Since the relation $\dot \Omega\propto \Omega^3$ does not generally hold for measured pulsar properties, Eq.~\ref{eq:omegadot3} is usually replaced 
with the more general form 
\begin{equation}
    \dot \Omega =-k\ \Omega^n
    \label{eq:omegadotn}
\end{equation}
with $n=\ddot \Omega \Omega / \dot \Omega^2$, the so-called {\it pulsar braking index}. The latter equation provides an operative way to estimate $n$ when measurements of the pulsar period are available on a sufficiently long time-span. In general one finds $n\neq 3$.
Looking at Eq.~\ref{eq:omegadot3}, without invoking non-electromagnetic spin-down processes, one can interpret this result in terms of variation with time of the quantities defining the relation between $\Omega$ and $\dot \Omega$. More specifically:
\begin{equation}
n=3+\frac{\Omega}{\dot \Omega} \left(2 \frac{\dot B_\star}{B_\star}+2 \dot \chi \frac{\cos\chi}{\sin\chi}+6 \frac{\dot R_\star}{R_\star}-\frac{\dot I_\star}{I_\star}\right)\ .
    \label{eq:nvar}
\end{equation}

Eq.~\ref{eq:omegadotn} can be integrated in time, leading to:
\begin{equation}
    \Omega(t)=\frac{\Omega_0}{(1+t/\tau)^{1/(n-1)}
    \label{eq:omegat}}
\end{equation}
and for the spin-down power:
\begin{equation}
\dot E(t)=\frac{\dot E_0}{(1+t/\tau)^\alpha}
\label{eq:edott}
\end{equation}
with $\alpha=(n+1)/(n-1)$ and $\tau=P_0/((n-1)\dot P_0)$. $\tau\approx t_{ch}$ is often used as an estimate for $\tau$, while $t_{ch}$, $B_{\rm min}$ and $\dot E$ (and $n$ when sufficient data are available) are all derived from timing analysis. For the specific case of e.m. spin-down , one has 
\be
\tau=\frac{3 c^3 I_\star P_0^2}{4 \pi^2 B_\star^2 R_\star^6}=2 \times 10^{15} {\rm s}\ I_{45} B_{12}^{-2}\ R_{\star,6}^{-6} P_{1}^2
\label{eq:tausd}
\ee

While pulsed emission was at the origin of pulsar discovery, it stays as one of the most mysterious aspects of pulsar physics: the mechanisms behind it at different energies are still debated and we will not discuss them at all in these lectures. For our purposes, however, it is important to notice that in terms of global energetics, the energy that goes into pulsed radiation is a relatively small fraction ($\sim 10^{-6}$ at radio wavelengths and of order tens of \% in gamma-rays) of the total energy lost by the pulsar per unit time. As we will see, most of the rotational energy a pulsar loses is emitted instead in the form of a magnetized relativistic wind, which then gives origin to a Pulsar Wind Nebula. In the following we briefly discuss the origin of this wind.

\subsection{The Goldreich and Julian (1969) magnetosphere}
\begin{figure}[h!!!]
\begin{center}
\includegraphics[width=0.35\textwidth]{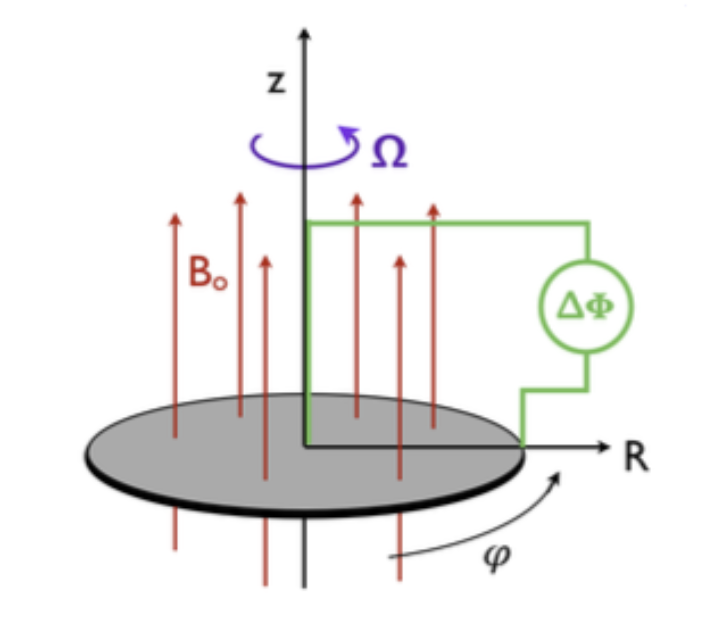}
\hspace{1cm}
\includegraphics[width=0.55\textwidth]{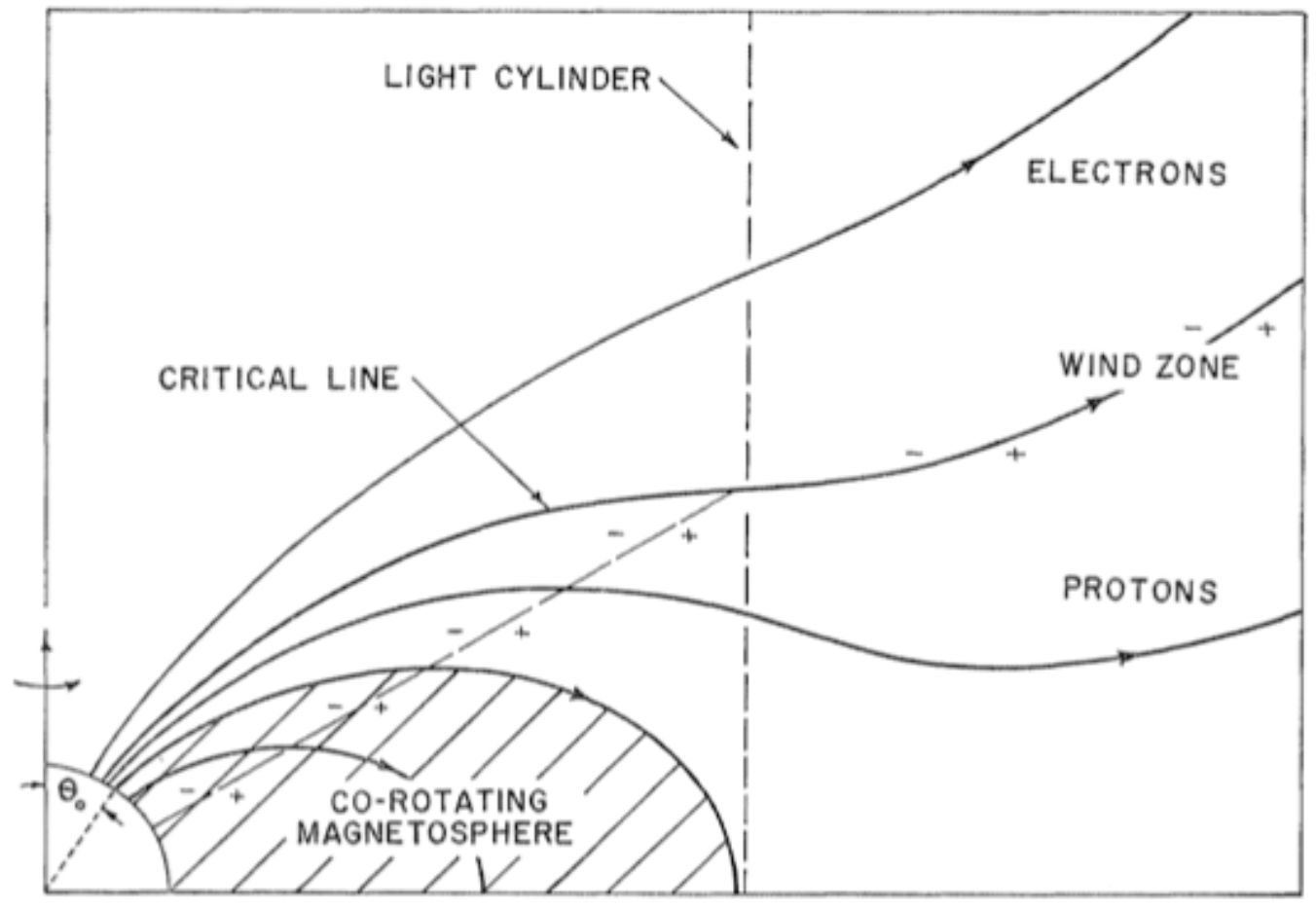}  
\caption{\footnotesize{Left panel: a Faraday disk. Right panel: the Goldreich and Julian Magnetosphere, taken from \cite{GJ69}.}}
\label{fig:gjpsr}
\end{center}
\end{figure}

The basic picture of how a pulsar would release particles and form a wind was proposed by \cite{GJ69}, who were the first to realize that a pulsar would surround itself by a magnetosphere similar to the one in the right panel of Fig.~\ref{fig:gjpsr}.

Let us consider the simplest case of an aligned rotator, namely a pulsar whose dipole magnetic moment is aligned with its rotation axis. The basic physics of this object is extremely similar, aside from geometrical complications, to that of a Faraday disk, namely a rotating conducting disk immersed in a uniform magnetic field parallel to its rotation axis (left panel of Fig.~\ref{fig:gjpsr}). The free charges within the disk will try to set-up a force-free equilibrium, namely they will arrange so as to cancel the total force. This requires compensating the Lorentz force acting on the charges with the generation of an electric field $\vec E=-\frac{\Omega r}{c} B_0 {\bf e_r}$, directed along the disk radius and creates a potential difference $\Delta \Phi_d$ between the disk axis and its radius $R_d$:
\begin{equation}
\Delta \Phi_d=\int_0^{R_d} \vec E \cdot d\vec r=\frac{\Omega R_d^2 B_0}{2c}
\label{eq:dphidisk}
\end{equation}
The charge density in the disk can simply be computed as:
\begin{equation}
\rho_{disk}=\frac{\vec \nabla \cdot \vec E}{4 \pi}=\frac{1}{4 \pi r}\frac{d}{dr} (rE_r)=-\frac{\Omega B_0}{2 \pi c}=\frac{\vec \Omega \cdot \vec B_0}{2 \pi c} .
\label{eq:rhodisk}
\end{equation}
If we now move to the pulsar, the basic physics is unchanged. Assuming that the star is a perfect conductor, the charges within the star, co-rotating with velocity 
\be
\vec v=\frac{\Omega R \sin \theta}{c} {\bf e}_\phi\ ,
\label{eq:corot}
\ee
where ${\bf e}_\phi$ is the azimuthal unit vector in a spherical coordinate system $({\bf e}_R,{\bf e}_\theta, {\bf e}_\phi)$, will arrange in such a way as to guarantee
\be
\vec E^{in}=-\frac{\vec v \wedge \vec B}{c}=\frac{B_\star}{2} \left(\frac{R_\star}{R}\right)^3 \frac{\Omega R}{c} \left[ \sin^2\theta {\bf e}_R-2 \sin \theta \cos \theta {\bf e}_\theta\right]
\label{eq:efieldin}
\ee
everywhere within the star.

If we assume that there are no charges outside the star, then we can also compute the field outside, by solving the Poisson equation for the electric potential in vacuum, $\nabla ^2 \Phi=0$, with the boundary condition that the tangential component of the electric field at the star $E_\theta^{out}=-(1/R)(\partial \Phi/\partial R)$ matches $E_\theta^{in}$ given by Eq.\ref{eq:efieldin}. Once this is done, the result is:
\be
\Phi(R,\theta)=-\frac{B_\star \Omega R_\star^2}{6c} \left(\frac{R_\star}{R}\right)^3 (3 \cos^2 \theta-1)\ .
\label{eq:Phivac}
\ee
The condition $\Phi(R,\theta_c)=0$ identifies the so-called {\it critical line}, leaving the star surface at $\theta_c=\arccos\sqrt{1/3}$ (see right panel of Fig.\ref{fig:gjpsr}).

This solution is easily seen to be inconsistent: in fact, if one derives the radial component of the electric field $E_R^{out}$ from Eq.\ref{eq:Phivac}, this is found to differ at the star surface from $E_R^{in}$. A discontinuity in the electric field component perpendicular to the star surface implies that some charge surface density will accumulate there:
\be
\sigma_\star=\frac{E_R^{out}-E_R^{in}}{4 \pi}=-\frac{B_\star \Omega R_\star}{4 \pi c}\cos^2 \theta\ .
\label{eq:sigmax}
\ee
The next question that it is natural to ask is whether these negative\footnote{Note that the sign of the charge surface density depends on the relative orientation of $\vec \mu$ and $\vec \Omega$. $\sigma_\star$ would be positive everywhere if $\vec \Omega$ and $\vec B$ were counter-aligned, rather than aligned} charges can be kept on the star. Given that the radial component of the electric field $E_R(R_\star)$ is negative, electrons will be pulled out of the star along magnetic field lines, unless the gravitational force, $\vec F_g$, is strong enough to contrast the electric force. This is easily shown not to be the case, since 
\be
\left|\frac{e \vec E \cdot \vec B}{\vec F_g \cdot \vec B}\right|=\frac{e\ B_\star^2 \Omega R_\star}{c}\frac{R_\star^2}{G M_\star m_e B_\star}\cos^2 \theta\approx 8 \times 10^{11} \frac{B_{12}\ R_6^3}{M_{\star,\odot}\ P_1}\cos^2 \theta
\label{eq:starforce}
\ee
where $R_6$ and $M_\star,\odot$ are the star radius and mass in units of $10^6$ cm and solar mass, respectively, $B_{12}$ is the star magnetic field in units of $10^{12}$ G and $P_1$ is the star period in seconds. 

Eq.\ref{eq:starforce} demonstrates that electrons (and the same holds true for protons in the case of a counter-aligned rotator) are easily pulled out of the star in principle, so that \cite{GJ69} concluded that the star surrounds itself of a co-rotating magnetosphere, as shown in the left panel of Fig.\ref{fig:gjpsr}. 

Co-rotation implies an increasing speed of the particles with increasing distance (see Eq.\ref{eq:corot}) and can only extend out to the so-called {\it light cylinder radius} $R_L=c/\Omega$, beyond which the required speed would exceed the speed of light. As one can see from the right panel of Fig.\ref{fig:gjpsr}, some of the dipole field lines extend beyond $R_L$ and can in principle take the particles indefinitely far from the star. These {\it open field lines} are identified by the condition that they intercept the star surface at a latitude that is less than $\theta_{pc}$ with $\theta_{pc}$ identifying the foot of the last closed field line, namely the dipole line that is tangent to the light cylinder at the equator. Using the dipole field line equation, $dR/R d\theta=B_R/B_\theta=2 \cos \theta/\sin\theta$, leading to $R(\theta)=R_0 \sin^2 \theta$, and the condition that $R(\pi/2)=R_0=R_L$, one readily finds 
\be
\sin^2\theta_{pc}=R_\star/R_L .
\label{eq:thpc}
\ee

Particles leaving the star along the open field lines can reach infinity with very high energy. Let us now compute the potential that is available for their acceleration. From Eq.\ref{eq:Phivac}, we can compute the full potential drop between the star pole and infinity as
$\Phi_\infty=\Phi(R_\star,0)-\Phi(R_\star,\theta_c)$, obtaining:
\be
\Delta\Phi_\infty=\frac{B_\star \Omega R_\star^2}{2c}\approx 3 \times 10^{16}\ B_{12}\ R_6^2\ P_1^{-1} {\rm V}\ .
\label{eq:phiinf}
\ee
Not surprisingly, this is exactly the same expression as for the Faraday disk (Eq.\ref{eq:dphidisk}) and for typical pulsar parameters the full potential difference is huge. The potential difference $\Delta\Phi_\infty$, however, is not fully available to the particles that leave the system. These can only experience the fraction of this drop that develops between the star pole and the last closed field line. From Eq.\ref{eq:Phivac}, the result is readily found to be:
\be
\Delta\Phi_{pc}=\frac{B_\star \Omega R_\star^2}{2c} \frac{R_\star}{R_L}\ ,
\label{eq:psrdrop}
\ee
namely a fraction $R_\star/R_L=2 \times 10^{-4} R_6 /P_1$ of the total available drop, increasing for short rotation periods. A possible way to make a larger fraction of $\Delta \Phi_\infty$ available to the particles that leave the magnetosphere is to open a larger fraction of field lines, something that might perhaps happen occasionally, due to instabilities, but as we will better motivate later on, cannot involve a large amount of particles  over long time-scales without reflecting in a change of the pulsar bolometric spin-down luminosity.

As we will see in the following $\Phi_{pc}$ is the most important quantity to define the maximum particle energy that can be achieved in a pulsar powered system, not only the pulsar, but even the PWN. It is therefore important to understand how well it can be determined from pulsar observations.

Another important quantity to assess is the flux of particles that leave the star magnetosphere. The charge density required to ensure that the magnetosphere reaches a steady-state configuration with no unscreened electric field along $\vec B$ can be worked out as $\rho_e=\vec \nabla \cdot \vec E/(4 \pi)$, with $\vec E$ the corotation electric field. The result is the so-called {\it Goldreich \& Julian} charge density \cite{GJ69}, usually written as\footnote{With respect to the actual derivation by \cite{GJ69}, a geometrical factor related to the dipolar field structure, that is important near the star, is here neglected}:
\be
\rho_{GJ}=-\frac{\vec \Omega \cdot \vec B}{2\pi c} ,
\label{eq:rhogj}
\ee
which is again essentially the same as what we found for the Faraday disk (Eq.\ref{eq:rhodisk}). It is to be noted that $\rho_{GJ}$ changes sign at $\theta_c$, where also $\Phi(\theta_c)=0$.

If the only source of particles in the magnetosphere were the star surface, the flux of particles released in the pulsar surrounding would be constrained to be the integral over the polar cap of a density $\rho_{GJ}/e$ of particles leaving the star at the speed of light: 
\be
\dot N_{GJ}=\frac{B_\star \Omega^2 R_\star^3}{2\ e\ c}\ .
\label{eq:ndotGJ}
\ee

\subsection{The pulsar wind}
\label{sec:psrwind}
\begin{figure}[h!!!]
\begin{center}
\includegraphics[width=0.5\textwidth]{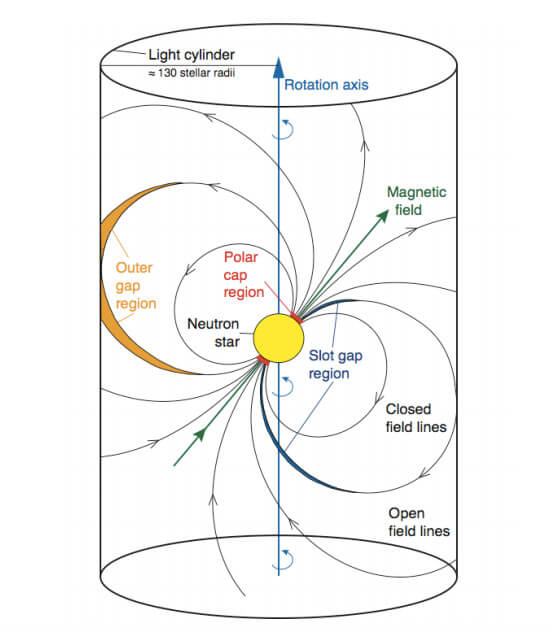}  
\hspace{1cm}
\includegraphics[width=0.4\textwidth]{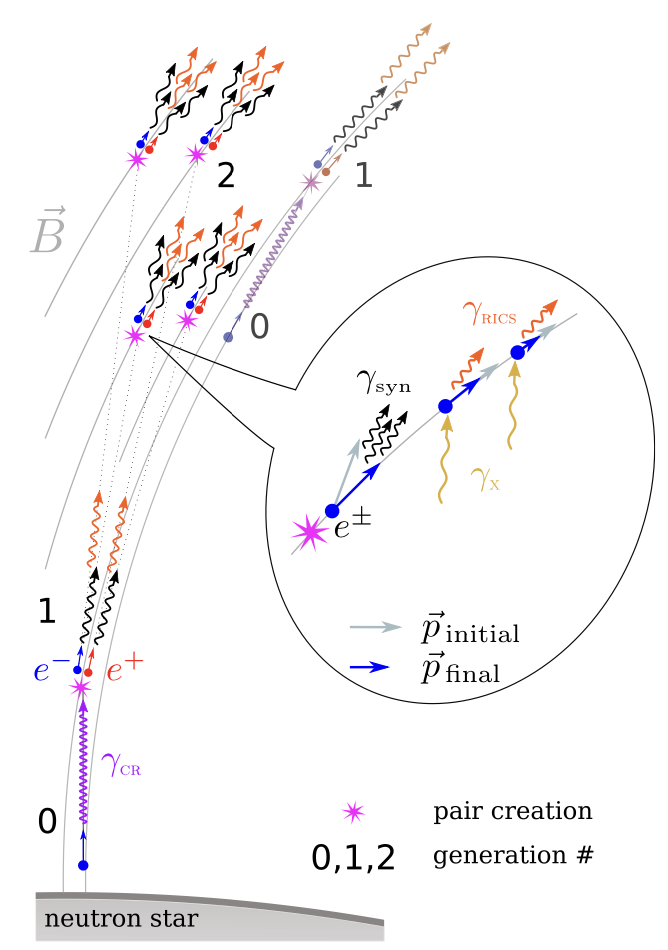}
\caption{\footnotesize{Left panel: pair creation regions in the pulsar magnetosphere (Credit: MAGIC Collaboration). Right panel: Illustration of the pair production process (figure from \cite{TimokhinHarding19}).}}
\label{fig:pairprod}
\end{center}
\end{figure}

In fact, while this estimate is solid in terms of providing the flux of protons, or nuclei\footnote{remember that the neutron star surface is likely made of a {\it Fe} lattice} that leave the magnetosphere, it only constitutes a wild lower limit for the flux of electrons and positrons. Very efficient pair-production must take place in the pulsar magnetosphere, as derived both from theoretical considerations and observations. In terms of theory, a closer look at the magnetosphere (Fig.\ref{fig:gjpsr}) immediately shows that $\rho_{GJ}$, the required charge density to guarantee the condition $\vec E \cdot \vec B=0$, cannot be ensured everywhere if the only charges available are those emitted from the star. The main regions in which this condition is difficult to guarantee are illustrated in the left panel of Fig.\ref{fig:pairprod}. These are: 1) the polar cap regions at the star surface, in case charges of the right sign cannot be extracted from the star; 2) the slot gaps, still in the polar cap region but at somewhat higher altitude, where depending on the inclination between $\vec \Omega$ and $\vec B$ star supplied charges might not be enough as the distance from the star increases (Space Charge Limited Flow, \cite{AronsSCLF}); 3) the outer gaps, where the last closed field line intersects the critical line at $\theta_c$, where the sign of $\rho_{GJ}$ changes \cite{Ruderman75}. 

We will not discuss pair production in any detail in these lectures (a recommended recent article on the subject, also rich of references to the classics, is \cite{TimokhinHarding19}) but the basic physics picture, also illustrated in Fig.\ref{fig:pairprod}, is as follows: in regions of unscreened electric field parallel to $\vec B$, any charge is bound to accelerate very quickly, and hence radiates (e.g. by synchrotron or curvature radiation); the emitted photons are above threshold for pair production in the intense ambient magnetic field and the produced pairs can emit pair-producing photons in turn. A pair-production cascade develops and goes on until a sufficient number of charges to screen the field is produced. This process can easily fill the magnetosphere with a number of charges that is much larger than $\rho_{GJ}/e$. The ratio between the total flux of charges leaving the magnetosphere and $\dot N_{GJ}$ is the so-called {\it pulsar multiplicity}, $\kappa$. The most recent estimates of $\kappa$ for young pulsars are in the range $10^3-10^5$ \cite{TimokhinHarding19}. These values ensure that electron-positron pairs are, by number, largely dominant over ions in the pulsar outflow, since the latter, even if present, can be injected at most at a rate $\dot N_{GJ}$. It is important to notice, however, that, if all particles become part of a cold MHD wind, i.e. move with the same bulk Lorentz factor, much larger than any thermal or non-thermal dispersion, values of $\kappa$ towards the lower end of the estimated range leave the possibility open that ions might dominate over pairs in terms of energy.

All these considerations take us to the question of how these particles leave the magnetosphere.

From the point of view of observations, the privileged channel to investigate pair creation is $\gamma$-rays Astronomy, since the pair-creating radiation falls by definition in the gamma-ray range, and interesting constraints have come, in recent years, both from high-energy ($>0.5$ MeV) \cite{Abdo13} and VHE ($>0.1$ TeV) emission \cite{CrabMAGICTeV}. The results of these recent high energy observations suggest that pair production can partly occur at large distance from the pulsar, outside the light cylinder.

In order to understand how this is possible and to connect with the physics of PWNe, we need to understand the pulsar environment at scales larger than the light cylinder. The self-consistent dynamical modeling of the pulsar magnetosphere has been a formidable challenge since the discovery of these sources. We mentioned that the early Goldreich \& Julian model of the magnetosphere of an aligned rotator is not self-consistent, but in fact, a satisfactory description even of this simplest case had to wait about 30 years, when a solution for the steady-state force-free problem\footnote{force-free means that the particle inertia is set to zero, an assumption that is valid as long as the dynamics is fully dominated by the electromagnetic field} could be found numerically \cite{Contopoulos99}. Interestingly, this solution showed that at large distances compared to the light cylinder, the outflow from an aligned magnetic dipole is essentially the same as for an aligned split monopole, namely a monopole, and hence fully radial, magnetic field, with opposite directions in the two hemispheres. The solution of the latter problem is analytical and known since the early '70s \cite{Michel74}. The velocity expands radially ad the speed of light and the electric and magnetic field beyond $R_L$ can be written as:
\be
\left\{
\begin{array}{ccl}
\vec B_{\rm far}&=&B_L \left[\left(\frac{R_L}{R}\right)^2 {\bf e}_R - \left(\frac{R_L}{R}\right) \sin \theta\ {\bf e}_\phi \right]\\
\\
\vec E_{\rm far} &=& - B_L \left(\frac{R_L}{R}\right) \sin \theta\ {\bf e}_\theta\
\end{array}
\right. .
\label{eq:split}
\ee

The expressions in Eq.\ref{eq:split} imply that the magnetic field is predominantly toroidal at large distances and the energy flux, which is all in the form of Poynting flux is:
\be
\vec S(R,\theta)=\frac{c B_L^2}{4 \pi}\left(\frac{R_L}{R}\right)^2\sin^2\theta\ {\bf e}_R,
\label{eq:eflux}
\ee
namely maximum at the rotational equator and decreasing towards the poles. The total energy lost by the pulsar per unit time can be computed as the integral of $S(R,\theta)$ on any spherical surface at distance $R$: $\dot E=(2/3)c\ B_L^2 R_L^2$. Taking into account that $B_L=B_\star (R_\star/R_L)^3$, the result is:
\be
\dot E=\frac{2}{3}\frac{B_\star^2 \Omega^4 R_\star^6}{c^3}\ .
\label{eq:edotgen}
\ee
This expression is the same, aside from factors of order unity, as what we had found for the power emitted by an oblique rotator (Eq.\ref{eq:edotdip}), but it applies to a system that would not emit any electromagnetic radiation based on that formula. In fact, the self-consistent description of the plasma around the pulsar implies that most of the energy lost by the pulsar does not go in direct e.m. emission, but rather in the launching of a magnetized relativistic wind. 

\begin{figure}[h!!!]
\begin{center}
\includegraphics[width=0.4\textwidth]{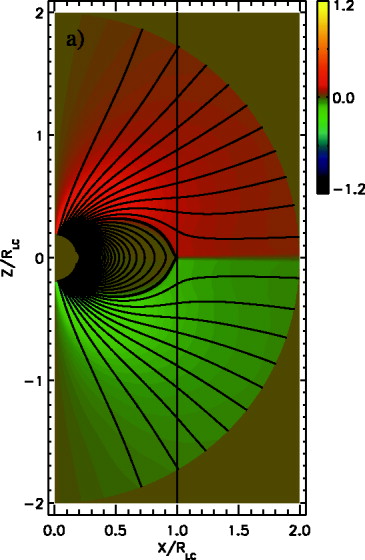}  
\hspace{1cm}
\includegraphics[width=0.48\textwidth]{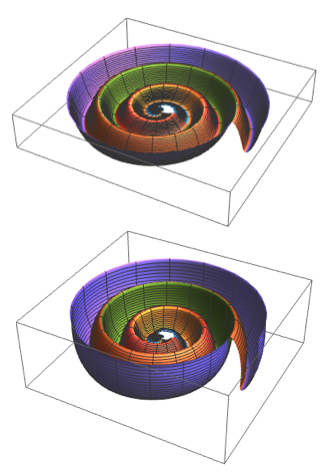}
\caption{\footnotesize{Left panel: the magnetosphere of an aligned dipole. The color scale represents the toroidal magnetic field superimposed on the poloidal field lines (picture from \cite{Spitkovsky06}). The toroidal field is null at the equator. Right panel: the current sheet that develops around the equatorial plane for an oblique rotator with $\chi=10^\circ$ (top) and $\chi=20^\circ$ (bottom).}}
\label{fig:FFPSR}
\end{center}
\end{figure}
Nowadays, full time-dependent solutions both in force-free \cite{Spitkovsky06}, MHD and full kinetic theory have become available for arbitrary inclinations, and all of them confirm this picture. The current attempts aim at taking into account also pair-production in a self-consistent manner. Recent and comprehensive reviews can be found in \cite{CeruttiBelo17,PhilippovKramer}. Fig.\ref{fig:FFPSR} illustrates the results of this modeling. The left panel of the figure shows the force-free magnetosphere of an aligned rotator ($\chi=0^\circ$), with the toroidal magnetic field scaling with latitude as $\sin\theta$ and changing sign in the equatorial plane. In the case of an oblique rotator, the toroidal field structure is more complex: on top of the $\sin\theta$ dependence, a region develops in which lines of opposite polarity alternate. This is the so-called {\it striped wind region}, whose angular extent depends on the pulsar obliquity. The left panel of Fig.\ref{fig:FFPSR} shows the equatorial current sheet between the alternating field lines for two different values of the obliquity ($\chi=10^\circ$ on top and $\chi=20^\circ$ at the bottom). 

The presence of alternating field lines makes the striped wind region an ideal location for magnetic reconnection to take place. It is interesting to notice that if reconnection took place in the freely-expanding wind region, and led to particle acceleration, this would likely result in pulsed (rather than steady-state) gamma-ray emission \cite{KirkGallant02}, granted that the wind Lorentz factor is large enough: such a scenario is currently being considered in light of the VHE gamma-ray results \cite{CrabMAGICTeV} that move the origin of gamma-ray pulses, and also some of the pair production process \cite{PhilippovKramer}, progressively further away from the pulsar.  

A way to dissipate magnetic energy between the pulsar light cylinder and the PWN has long been looked for by pulsar and PWN theorists. The reason for this is that while at the light cylinder the wind must be largely Poynting flux dominated (this is an electromagnetically driven outflow), modeling of PWNe suggests that when the outflow reaches asymptotic conditions the amount of energy carried by the particles has increased to the point to equal or even surpass the amount of energy carried by the magnetic field. The relative importance of magnetic fields and particle kinetic energy in the wind is summarized by the parameter $\sigma$ defined as the ratio between Poynting flux and particle kinetic energy:
\be
\sigma=\frac{B^2}{4 \pi n_\pm m_e \Gamma_w^2 c^2}
\label{eq:sigma}
\ee
where $m_e$ is the electron mass, $n_\pm$ the electron (and positron) density in the proper frame and $\Gamma_w$ is the Lorentz factor of the flow. At the light cylinder $\sigma$ must be large ($\sigma_{LC}\gg 1$) and $\Gamma_w$ only moderate (negligible particle inertia). If beyond the light cylinder the wind expands according to ideal MHD $\sigma$ will stay large, but this is in contrast with observations of PWNe, that indicate, for the asymptotic wind properties $10^3\lesssim\Gamma_w \lesssim 10^7$ and $0.1\lesssim \sigma \lesssim 10$.

Even without detailed knowledge of how the energy of the outflow is shared between the different components, what should be clear by now is that the wind is the primary channel through which the star loses energy, so that its total luminosity must equal $\dot E$. Hence:
\be
\dot E= \kappa \dot N_{GJ} m_e \Gamma_w c^2 \left(1+\frac{m_i}{\kappa m_e}+\sigma\right)\ ,
\label{eq:winden}
\ee
where $m_i$ is the ion mass. This expression clearly shows that if $\kappa<m_i/m_e$ (as can be the case for $\kappa\lesssim 10^4$) ions can be energetically dominant over pairs in the outflow, in spite of being numerically a small fraction. 

\subsection{Pulsars as UHECR sources}
\label{sec:uhecrs}

When discussing pulsars and PWNe as CR sources, the obvious questions to ask is what composition and what maximum energy these sources can provide. We had already found an expression for the maximum available potential drop (Eq.\ref{eq:psrdrop}), but now that we have found that Eq.\ref{eq:edotgen} is generally valid to describe the power emitted by a pulsar, we can use it to reformulate some of the quantities we had earlier expressed in terms of pulsar parameters: $\Delta\Phi_{pc}$ (Eq.\ref{eq:psrdrop}), and $\dot N_{GJ}$ (Eq.\ref{eq:ndotGJ}) will now read (neglecting factors of order unity):
\be
    \Delta \Phi_{pc}=\sqrt{\frac{\dot E}{c}}. \ \ \ \ \ \ \ \ \ \ \dot N_{GJ}=\frac{\sqrt{c\ \dot E}}{e}
\label{eq:dphingj}
\ee
From the expressions above it is clear that the energy lost by the pulsar can also be interpreted as carried away by a flux of particle $\dot N_{GJ}$ leaving the star with energy equal to the maximum available drop: another possible way to look at $\dot E$.

It is also clear that both quantities are immediately known once $\dot E$ is measured for a given pulsar (a measurement of $P$ and $\dot P$ is sufficient, without any further assumption). The relations in Eq.\ref{eq:dphingj} also clarify the previous statement that it is not possible to steadily accelerate a large fraction of particles beyond $\Delta\Phi_{pc}$ without violating the overall energetics of the system.

\begin{figure}[h!!!]
\begin{center}
\includegraphics[width=0.8\textwidth]{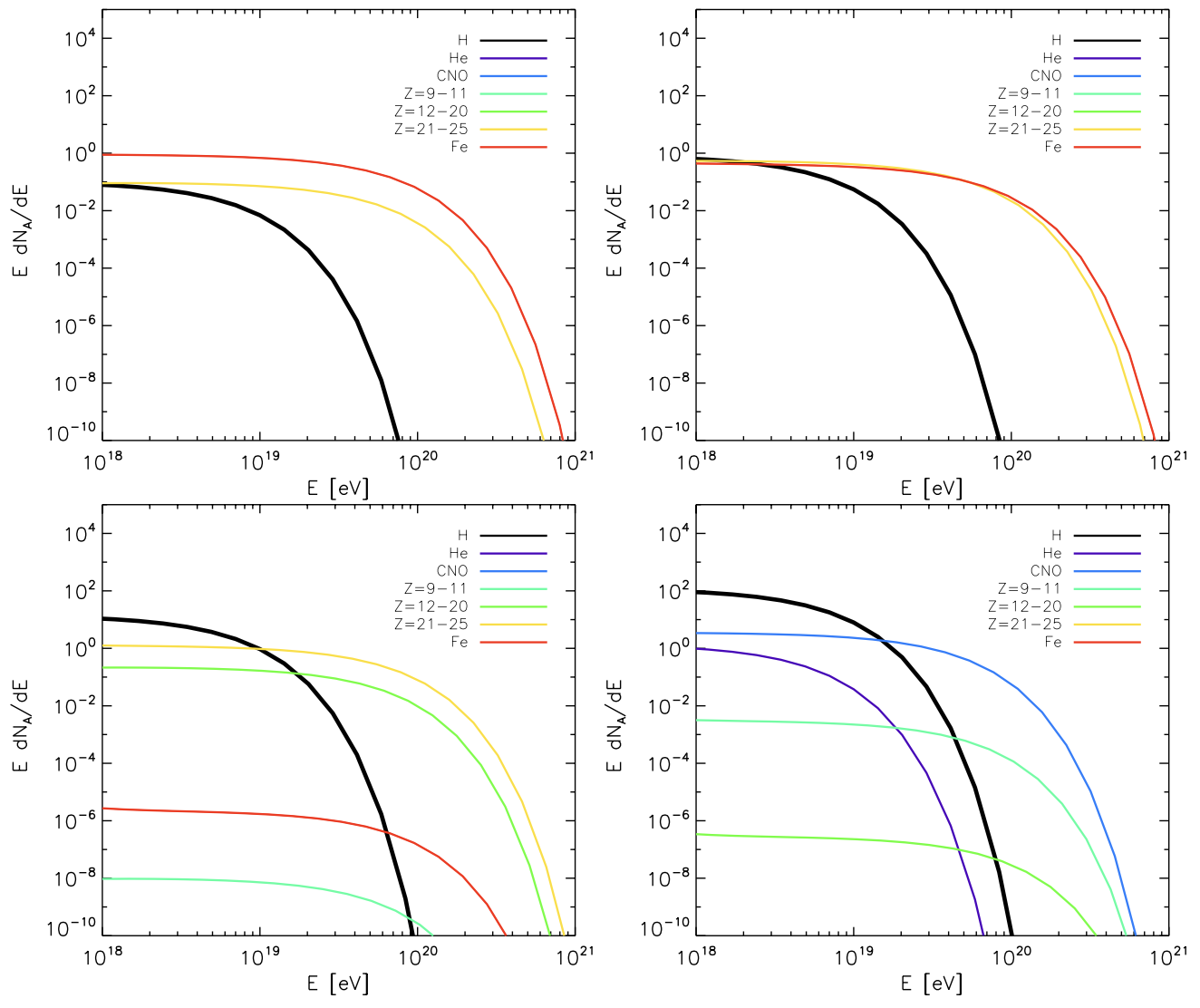}  
\caption{\footnotesize{Figure from \cite{Kotera15}. 
Cosmic-ray spectra $EdN_A/dE$ and composition produced by a highly magnetized pulsar with $B\approx 10^{13}$ G. The estimate refers to time $t_w$ after injection with $t_w = 100 \times R_L/c$ and an injection time $<100 yr$. The results refer to a case in which only iron is extracted from the star surface and photo disintegration is taken into account as described in section 3. From left to right, top to bottom, temperatures are $T = [1, 2, 5, 10] \times 10^6$ K.
  }}
\label{fig:UHECR}.
\end{center}
\end{figure}
It is to be noted, however, that even the potential difference $\Delta \Phi_{pc}$ can be very large, especially for new born (and hence rapidly rotating) highly magnetized neutron stars. For example, a ms period pulsar with a magnetic field of $\sim 10^{13}$G can, in principle, accelerate particles to $E_{\phi}=3 \times 10^{20}$eV $Z B_{13}P_{-3}$ (with $Z$ the atomic number of the nucleus), hence in the Ultra High Energy Cosmic Ray (UHECR) range. In reality the particle energy will be limited by curvature losses and in fact the maximum achievable Lorentz factor will be $E_{max}=m_i c^2\ \times \min(\Gamma_w, \Gamma_\phi, \Gamma_{\rm curv})$, where $\Gamma_w$ is the wind Lorentz factor in Eq.\ref{eq:winden}, which depend on pulsar multiplicity and wind magnetization, $m_i c^2 \Gamma_\phi=Z e \sqrt{\dot E/c}$ and $\Gamma_{\rm curv}\approx 10^8$ for typical pulsar parameters, and mildly varying. The highest energies are obtained for $E_{max}=E_\phi=E_w$, which corresponds to a situation in which most of the wind energy flows as $\dot N_{GJ}$ ions with energy $E_{max}$ (a necessary condition is $\kappa<m_i/m_e$). In this case the particle spectrum released by the system over its lifetime as a UHCR accelerator is easily computed to be $N_{CR}(E)=\dot N_{GJ}(0) \tau E^{-1}$ where $\tau$ is the spin-down time in Eq.\ref{eq:tausd}.

Assuming that the particles initially extracted from the star surface are {\it Fe} nuclei\footnote{Current theories for the neutron star surface want it to be made of an Iron lattice}, the radiation from the star will cause photodisintegration during magnetospheric propagation, so that, depending on the star temperature, which determines the black body radiation field in its immediate vicinities, a mixture of different nuclei from $p$ to {\it Fe} will eventually be released in the ISM. As shown in Fig.\ref{fig:UHECR} (from \cite{Kotera15}) the resulting spectrum and composition can easily match that estimated for UHECRs \cite{AugerCompo}.   

\section{Lecture II: Pulsar Wind Nebulae}
\label{sec:pwnintro}
\begin{figure}[h!!!]
\begin{center}
\includegraphics[width=0.66\textwidth]{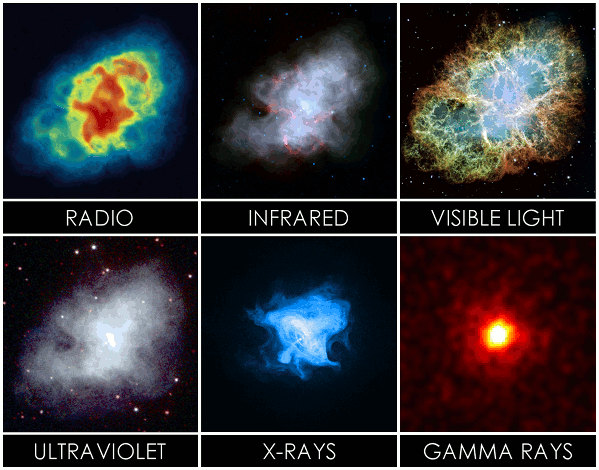}  
\hspace{.3cm}
\includegraphics[width=0.3\textwidth]{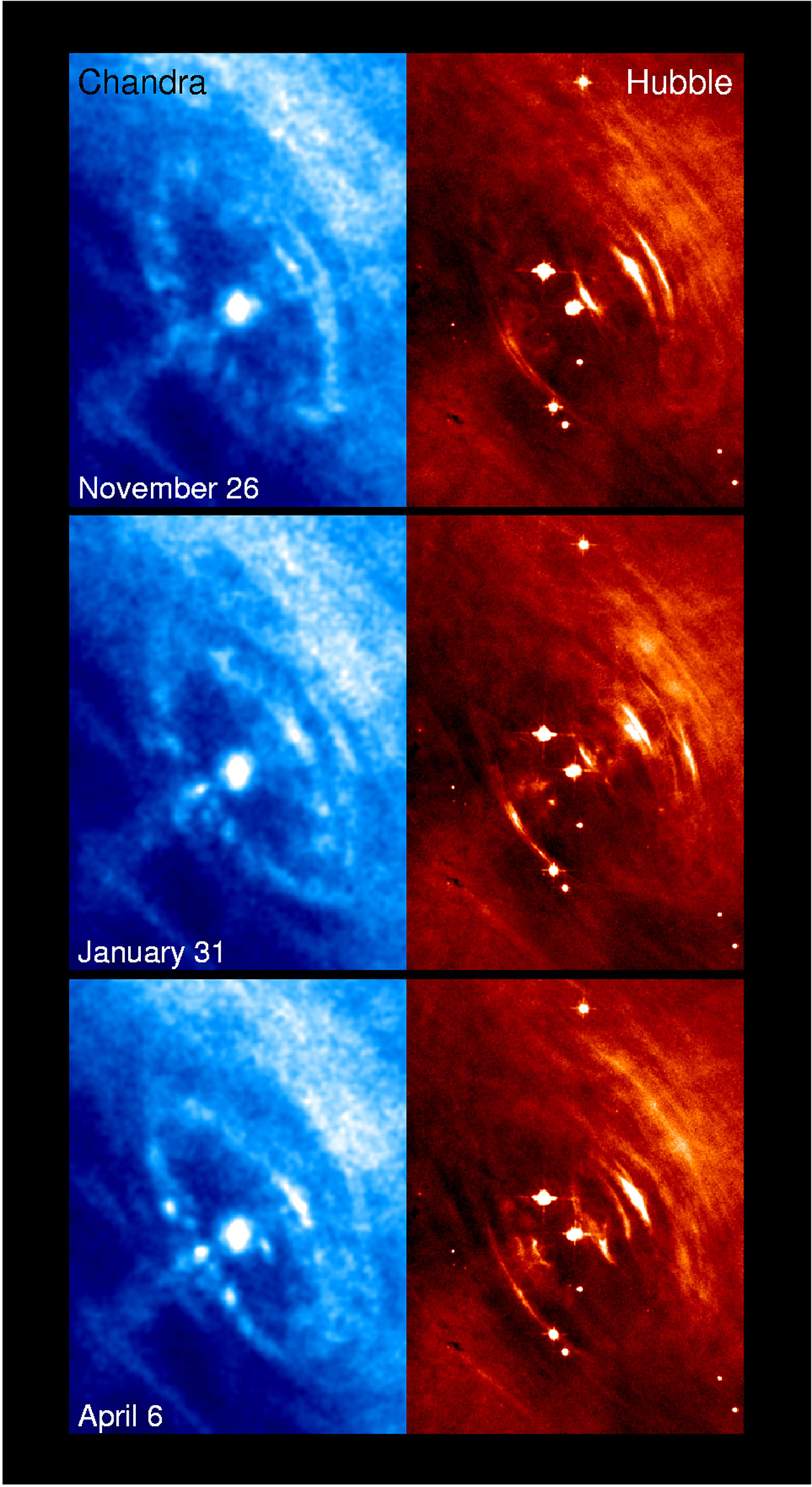}
\caption{\footnotesize{{\it Left panel:}the Crab Nebula observed in different wavebands, from radio to gamma-rays. {\it Right panel:} a zoom-in on the central region of the nebula at X-rays (left column by {\it Chandra} X-ray telescope) and optical (right column by {\it Hubble Space Telescope}) wavelengths. Snapshots at different times highlight the variability of the features immediately outside the central darker region.}}
\label{fig:crab}
\end{center}
\end{figure}

We are now ready to discuss what happens to the pulsar wind after it reaches its asymptotic structure and composition at large enough distance from the light cylinder. Before starting the discussion, it is appropriate to mention that we have learned most of what we know about PWNe from the class prototype, the Crab Nebula. This source, whose image is shown at difference frequencies in Fig.\ref{fig:crab} (spectrum in the top right panel of Fig.\ref{fig:sketch}), is one of the best studied objects in the Universe and its emission has been measured over more than 20 decades in frequency, from radio wavelengths to very high energy gamma-rays (see e.g.\cite{Universe21} for a recent review). The main radiation mechanism is synchrotron emission with a spectrum extending up to photon energies $\gtrsim$ 100 MeV, while at higher energies Inverse Compton scattering (ICS) becomes important.
One important thing to notice in Fig.\ref{fig:crab} is that the central region of the nebula is darker than its surroundings. This is consistent with the picture we drew earlier of a pulsar wind that leaves the outer magnetosphere as highly relativistic in terms of bulk motion but cold (hence non-radiative) and only becomes radiative after its kinetic energy is dissipated and turned into particle acceleration at a shock, easily identified with the highly dynamical region around the {\it Chandra} inner ring. The shock transition is caused by the fact that the wind must slow down in order to match the conditions of non-relativistic expansion of the surrounding SN ejecta, that act as a conducting cage. The expected position at which the wind termination shock (TS hereafter) must be located is easily worked out by approximating the PWN shape as spherical and assuming steady-state. Under these assumptions, the TS position will be determined as the distance from the pulsar at which the wind ram-pressure, $P_w=\dot E/(4 \pi c R^2)$ equals the pressure in the nebula, $P_N=\dot E t/(4/3 \pi R_N^3)$, where we have treated $\dot E$ as constant (a reasonable approximation for young systems). If we also approximate the nebular expansion speed $v_N$ as constant (again a reasonable approximation for young systems), so that the nebular radius is $R_N=v_N t$, the result is $R_{TS}=(v_N/c)^{1/2} R_N$. This position in the Crab Nebula (expanding since 1054 A.D. with an average speed of $\sim 1500$ km/s) coincides with the location where the {\it Chandra} inner ring and the highly dynamical wisps (right panel of Fig.\ref{fig:crab}) are observed. A sketch of a spherical PWN is reported in the left panel of Fig.\ref{fig:sketch}, where sizes in physical units refer to the Crab nebula case.

\subsection{One-zone models of PWNe}
\label{sec:onezone}
The first attempts at modeling PWNe and extracting the properties of the plasma within them from their bright non-thermal emission were carried out within the one-zone description, which basically considers the time-evolution of an expanding spherical bubble whose inner structure is ignored. The nebular magnetic field strength and particle population are derived by solving the following system of equations:
\be
N(\epsilon,t)=\int_{t_i}^t dt_i\ Q[\epsilon_i(\epsilon,t; t_i), t_i]\ \frac{\partial \epsilon_i}{\partial \epsilon}
\label{eq:NE-onezone}
\ee
where $N(\epsilon,t)$ is the number of particles per unit energy interval with energy between $\epsilon$ and $\epsilon+d\epsilon$ at the current time $t$ and $Q(\epsilon_i,t_i)$ is the injection rate of particles per unit energy interval with initial energy between $\epsilon_i$ and $\epsilon_i+d\epsilon_i$ injected at time $t_i$. The injection rate $J$ is normalized in such a way that a fraction $\eta_p$ of the pulsar wind energy is input in particles:
\be
\int d\epsilon_i\ \epsilon_i\ Q(\epsilon_i, t_i) =\eta_p \dot E(t)
\label{eq:jnorm}
\ee
and the relation between $\epsilon_i$ and $\epsilon$ is found by solving the evolution equation for the energy of a single particle subject to adiabatic and synchrotron losses:
\be
\frac{d\epsilon}{dt}=-\frac{\epsilon}{3R_N(t)} \frac{dR_N}{dt}-\frac{\sigma_T c}{9 \pi (m_e c^2)^2} B_N(t)^2 \epsilon^2
\label{eq:partevol}
\ee
where $\sigma_T$ is the Thompson cross section, $R_N(t)$ is the nebular radius as a function of time and in writing synchrotron losses we have assumed that the magnetic field in the nebula is isotropic and the field component perpendicular to the particle trajectory is $B_\perp=\sqrt{2/3} B_N$ with $B_N(t)$ the average magnetic field strength in the nebula. The last needed ingredient to close the system is the evolution law for the magnetic field. Assuming that a fraction $\eta_B$ of $\dot E$ goes into magnetic energy $W_B=B_N^2 R_N^3/6$ and that the field behaves as a relativistic fluid subject to adiabatic losses, one has:
\be
\frac{dW_B}{dt}=\eta_B \dot E-\frac{W_B}{3R_N}\frac{dR_N}{dt}.
\label{eq:Bevol}
\ee
A few things can be immediately learned by solving this set of equations for the Crab Nebula and other well surveyed PWNe \cite{NicJon11,Torres14}: 1)the magnetic field strength in PWNe, as derived from the ratio between the ICS and synchrotron component of the spectrum, is in the range $10-100$ $mu$G; 2) the particle injection rate is much larger than $\dot N_{GJ}$ with implied multiplicities in the range $few \times 10^4-10^6$; 3) the injected particle spectrum is a broken power-law:
\be
Q(E)=Q_0\times\left\{
\begin{array}{ccc}
 \left(\frac{E}{E_b}\right)^{-p_1} & E<E_b &\\
 \left(\frac{E}{E_b}\right)^{-p_2} & E>E_b &
 \end{array}
 \right.
\label{eq:spec}
\ee
with $p_1\approx 1.5$. $p_2\approx 2.3$ and $E_b\approx 500$ GeV. A similar spectrum of the particle population is derived also for other PWNe with sufficient spectral coverage \cite{Torres14,NicJon11}: one generally finds $E_b$ in the range 0.3-1 TeV, $1<p_1<2$ and $p_2\sim 2.2-2.5$.
\begin{figure}[h!!!!]
 \begin{center}
 \includegraphics[width=.47\textwidth]{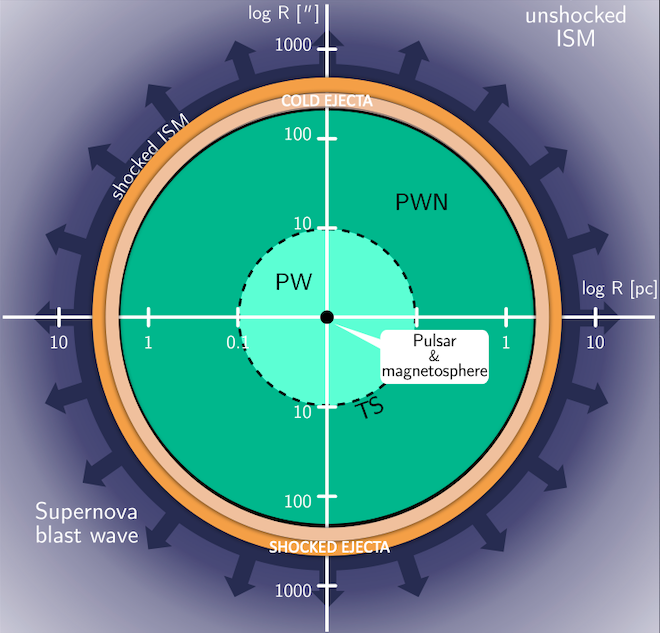} 
 \hspace{1cm}
 \includegraphics[width=.42\textwidth]{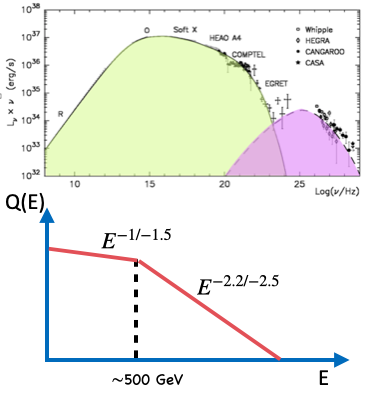}
 \caption{\footnotesize{
 {\it Left panel}: sketch of a young PWN, whose parent pulsar is still close to the center of the SN explosion (adapted from \cite{OlmiLuca17}). The black dot at the center is the pulsar, surrounded by its cold relativistic wind (PW, in light green). The wind is slowed down at the termination shock (TS, dashed circle), and gives rise to the PWN (dark green region). The latter is separated by a contact discontinuity from the cold SN ejecta (pink), through which the reverse shock of the SN blast wave propagates. Beyond the reverse shock and the heated up ejecta (orange), there is a shell of ISM material swept up by the SN forward shock (dark blue). Distances are in logarithmic scale, with numbers referring to the Crab Nebula \cite{KC1}. {\it Right panel}: on top is the emission spectrum of the Crab Nebula (adapted from \cite{Atoyan96}) due to synchrotron (green) and ICS emission (pink); the spectral shape of all well surveyed PWNe is similar and interpreted in terms of a typical population of injected particles as shown in the lower panel.   
  }}
\label{fig:sketch}
\end{center}
 \end{figure}

The schematic representation of a young system is shown in the left panel of Fig.\ref{fig:sketch}, while the right panel of the same figure shows the Crab Nebula emission spectrum and the underlying spectrum of accelerated particles at injection. It is to be noted that the high energy end of the synchrotron spectrum of the Crab Nebula is direct evidence of the presence in this source of PeV electrons (and positrons).

\subsection{The MHD description of PWNe}
label{sec:MHD}
The left panel of Fig.\ref{fig:sketch} is a re-elaboration of the original sketch of the Crab Nebula accompanying the first spatially resolved model of PWNe, the Kennel \& Coroniti model \cite{KC1}, namely a steady-state, spherically symmetric relativistic MHD description of the PWN dynamics and radiation properties. The basic set of equations to be solved is:
\begin{eqnarray}
\frac{1}{r^2}\frac{\partial}{\partial r} \left(r^2 n\ u \right)=0 
\label{eq:kcn}\\
\frac{1}{r^2}\frac{\partial}{\partial r}\left[r^2\left(n \mu u^2 +p+\frac{B^2}{8 \pi}\left(1+\frac{u^2}{\gamma^2} \right)\right)\right]-\frac{2p}{r}=0 \label{eq:kcmom}\\ 
\frac{1}{r^2}\frac{\partial}{\partial r}\left[r^2\left(n \mu \gamma u+\frac{B^2}{4 \pi}\frac{u}{\gamma}\right)\right]=0 \label{eq:kcen}\\
\frac{1}{r}\frac{\partial}{\partial r} \left(\frac{r u B}{\gamma}\right)=0 \label{eq:kcB}
\end{eqnarray}
where $n$ is the number density, $p$ is the pressure, $\mu=m c^2+(p/n)\Gamma_{ad}/(\Gamma_{ad}-1)$ is the relativistic enthalpy per particle (with $\Gamma_{ad}$ the plasma adiabatic index: $\Gamma_{ad}=4/3$ for a relativistic fluid), $B$ is the toroidal magnetic field (remember that we had found the toroidal component of the magnetic field, $B_\phi$ to be dominant in the wind beyond the light cylinder), all measured in the fluid rest-frame. Finally, $\gamma$ is the flow Lorentz factor and $u=\gamma \beta$ with $\beta=v/c$ and $v$ the flow speed. The equations Eq.(\ref{eq:kcn}-\ref{eq:kcB}) represent the conservation of number density, momentum and energy, and the induction equation.

If there are no discontinuities in the flow, after some manipulation, Eqs.(\ref{eq:kcn}-Eq.\ref{eq:kcB}) reduce to the following form:
\begin{eqnarray}
    \frac{\partial}{\partial r} \left( n u r^2\right)=0 \label{eq:kcn1}\\
    \frac{\partial}{\partial r} \left( \mu \gamma+\frac{B^2}{4 \pi n \gamma}\right) =0 \label{eq:kcmom1}\\
    \frac{\partial}{\partial r} \left( \frac{p}{n^{4/3}}\right)=0 \label{eq:kcen1}\\
    \frac{\partial}{\partial r} \left( \frac{ruB}{\gamma}\right)=0. \label{eq:kcB1}
\end{eqnarray}
When specializing to the case of a cold wind ($p\ll n m c^2$), namely $\mu\approx mc^2$, it is straightforward to see that Eq.\ref{eq:kcmom1} reduces to the conservation of energy per particle along a streamline:
$\gamma(1+\sigma)=const$, which makes it clear that, if at $R_{LC}$ the wind was magnetically dominated and with an energy flux $\propto \sin^2\theta$, we expect this angular dependence to persist in the following evolution, even if the wind accelerates and the ratio between kinetic and magnetic energy changes. This has important consequences as we will see in the following.

Further algebraic manipulation of Eqs.\ref{eq:kcn1}-\ref{eq:kcB1} leads to an expression for the wind terminal velocity ($v_\infty=v(r\to \infty)$) that turns out to be (see \cite{KC1}): 
\be
v_\infty=\frac{\sigma_0}{1+\sigma_0}c
\label{eq:vinfty}
\ee
where $\sigma_0$ is the magnetization at the wind basis.
This makes it clear that a high-$\sigma$ wind will tend to a relativistic asymptotic speed ($v_\infty \to c$, for $\sigma> 1$), while only a low-magnetization wind will slow-down: $v_\infty\to \sigma_0 c$ for $\sigma_0\ll 1$. As a consequence, in the case of the pulsar wind, the conditions of non-relativistic expansion of the SN ejecta can only be matched for $\sigma_0\approx v_N/c\ll 1$. In the case of the Crab Nebula, this implies that at the TS (where Eqs.(\ref{eq:kcn1}-\ref{eq:kcB1}) start to be valid after the non-adiabatic shock jump) one must have $\sigma_{TS}\approx v_N/c\approx 10^{-3}$.

The set of Eqs.(\ref{eq:kcn}-\ref{eq:kcB}) can be integrated around the TS to work out the shock jump conditions, that will read:
\begin{eqnarray}
n_2 u_2=n_1 u_1 \\
\mu_2 n_2 u_2^2 +p_2 +\frac{B_2^2}{8 \pi} \left(1+\frac{u_2^2}{\gamma_2^2}\right)=\mu_1 n_1 u_1^2 +p_1 +\frac{B_1^2}{8 \pi} \left(1+\frac{u_1^2}{\gamma_1^2}\right) \\
\mu_2 n_2 u_2 \gamma_2 +\frac{B_2^2}{4 \pi} \frac{u_2}{\gamma_2}=\mu_1 n_1 u_1 \gamma_1 +\frac{B_1^2}{4 \pi} \frac{u_1}{\gamma_1}\\
\frac{u_2 B_2}{\gamma_2}=\frac{u_1 B_1}{\gamma_1}    \ .
\end{eqnarray}
In the case of a cold relativistic wind $p_1\ll n_1 m c^2$ and $\gamma_1\approx u_1\gg 1$, after some lengthy algebra one can express the shock jump purely in terms of the flow-magnetization, finding:
\begin{eqnarray}
u_2^2=\frac{1}{2} \left[\frac{8 \sigma^2+10 \sigma+1}{8 (1+\sigma)}+\sqrt{\left(\frac{8 \sigma^2+10\sigma+1}{8(1+\sigma)}\right)^2-\frac{\sigma^2}{2(1+\sigma)}}\ \right]\\
 \frac{B_2}{B_1}=\frac{n_2 \gamma_2}{n_1 \gamma_1}=\frac{N_2}{N_1}=\frac{\gamma_2}{u_2}\\
 \frac{p_2}{n_1 m c^2 \gamma_1^2}=\frac{1}{4 u_2 \gamma_2} \left[1+\sigma\left(1-\frac{\gamma_2}{u_2}\right)-\frac{\gamma_2}{\gamma_1}\right]\ .
\end{eqnarray}
It is interesting to evaluate these apparently complicated expressions for the small and large $\sigma$ limit. The result is
\begin{eqnarray}
\lim_{\sigma\to \infty}u_2^2= \sigma \Rightarrow 
\left\{
\begin{array}{cc}
&\frac{p_2}{n_1 m c^2 \gamma_1^2}\to \frac{1}{8\sigma}\\ 
& \\
&\frac{B_2}{B_1}=\frac{N_2}{N_1}\to 1
\end{array}
\right.
\\
\\
\lim_{\sigma\to 0} u_2^2= \frac{1+9\sigma}{8} \Rightarrow
\left\{
\begin{array}{cc}
  &   \frac{p_2}{n_1 m c^2 \gamma_1^2}\to \frac{2}{3}\\ 
   &  \\
   & \frac{B_2}{B_1}=\frac{N_2}{N_1}\to 3
\end{array}
\right.
\end{eqnarray}
Once again, it is clear that efficient dissipation and deceleration of the wind is only possible for $\sigma\ll 1$, in which case 2/3 of the bulk kinetic energy of the flow are converted into plasma pressure (the largest fraction of which is in fact carried by non-thermal particles in the case of PWNe), while if $\sigma$ is large the flow does not slow down nor dissipates efficiently.

These findings are at the heart of the so-called $\sigma$ problem that has been central to the study of PWNe for several years. The conclusion that one derives from steady-state spherical MHD modeling is that $\sigma_{TS}\ll1$. Reconciling this requirement with the fact that at the light-cylinder $\sigma_{LC}\gg 1$ has proven a very hard problem in spite of the fact that these two scales are widely separated (in Crab $R_{TS}/R_{LC}\approx 10^9$). A change in $\sigma$ by several orders of magnitude is not possible within ideal MHD \cite{Michel69} and requires a very large density of the pulsar wind \cite{LyubKirk01} to be accomplished in the striped wind region, as originally proposed by \cite{Coroniti90}.

While still not completely solved, the $\sigma$ problem has been much alleviated by more refined modeling, overcoming the limitations of 1D steady-state MHD with multi-dimensional time-dependent numerical studies. The motivation to engage in this kind of studies was unrelated to the $\sigma$-problem and actually came from {\it Chandra} observations showing the wide-spread presence of a jet-torus structure in all spatially well-resolved PWNe. In particular it was mysterious how the jet could originate from a distance closer to the pulsar than the inferred position of the termination shock (see right panel of Fig.\ref{fig:crab}). Since collimation of the flow while still relativistic is inefficient \cite{Lyub02}, a possible suggestion was that the TS could be highly non-spherical and much closer to the pulsar along the jet axis as a result of the latitude dependence of the wind energy flux: by the same argument of pressure equilibrium that we used to estimate the TS position in the spherical case, one would now guess that if most of the energy flows in the equatorial plane of the pulsar rotation, the shock will be much closer to the star along the polar axis than at the equator.

A number of studies \cite{Komissarov:2003,Komissarov:2004,del-Zanna:2004,del-Zanna:2006, Volpi:2008,Olmi:2014} focused on axisymmetric MHD simulations taking into account the latitude dependence of the wind energy flux highlighted above. Simulations were typically performed assuming $F(r,\theta)\propto (1+\sigma_0 \sin^2\theta)$ with $\sigma_0$ a parameter quantifying the ratio between the equatorial and polar flux. In addition, effective magnetic reconnection in the striped wind region was assumed, so that the magnetic field immediately upstream of the shock was assumed as $B_\phi\propto \sqrt{\sigma_{TS}} \sin\theta G(\theta)$, with $G(\theta)\to 0$ as $\theta\to 0$.

\begin{figure}[h!!!!]
 \begin{center}
 \includegraphics[width=.35\textwidth]{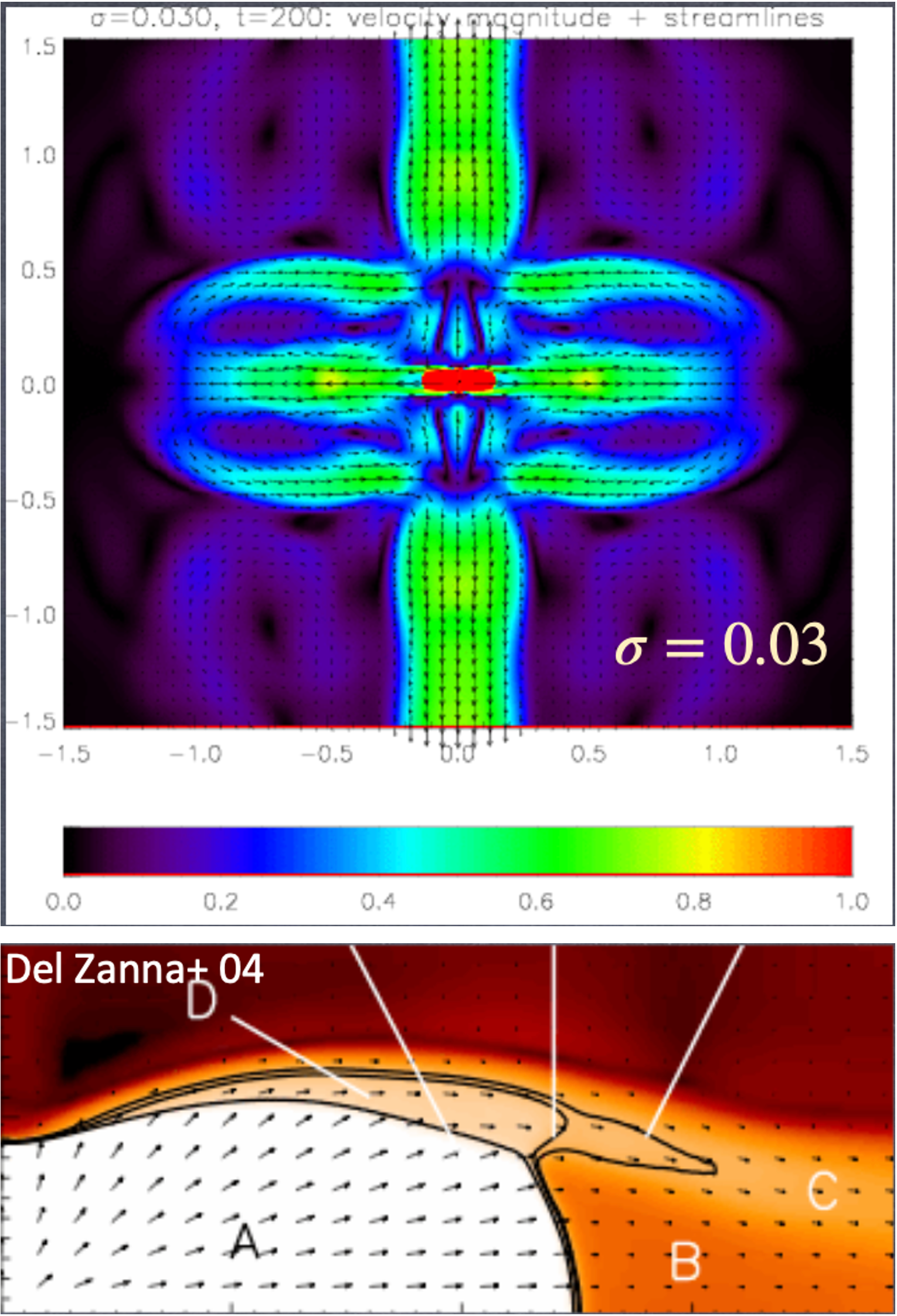} 
 \hspace{1cm}
 \includegraphics[width=.55\textwidth]{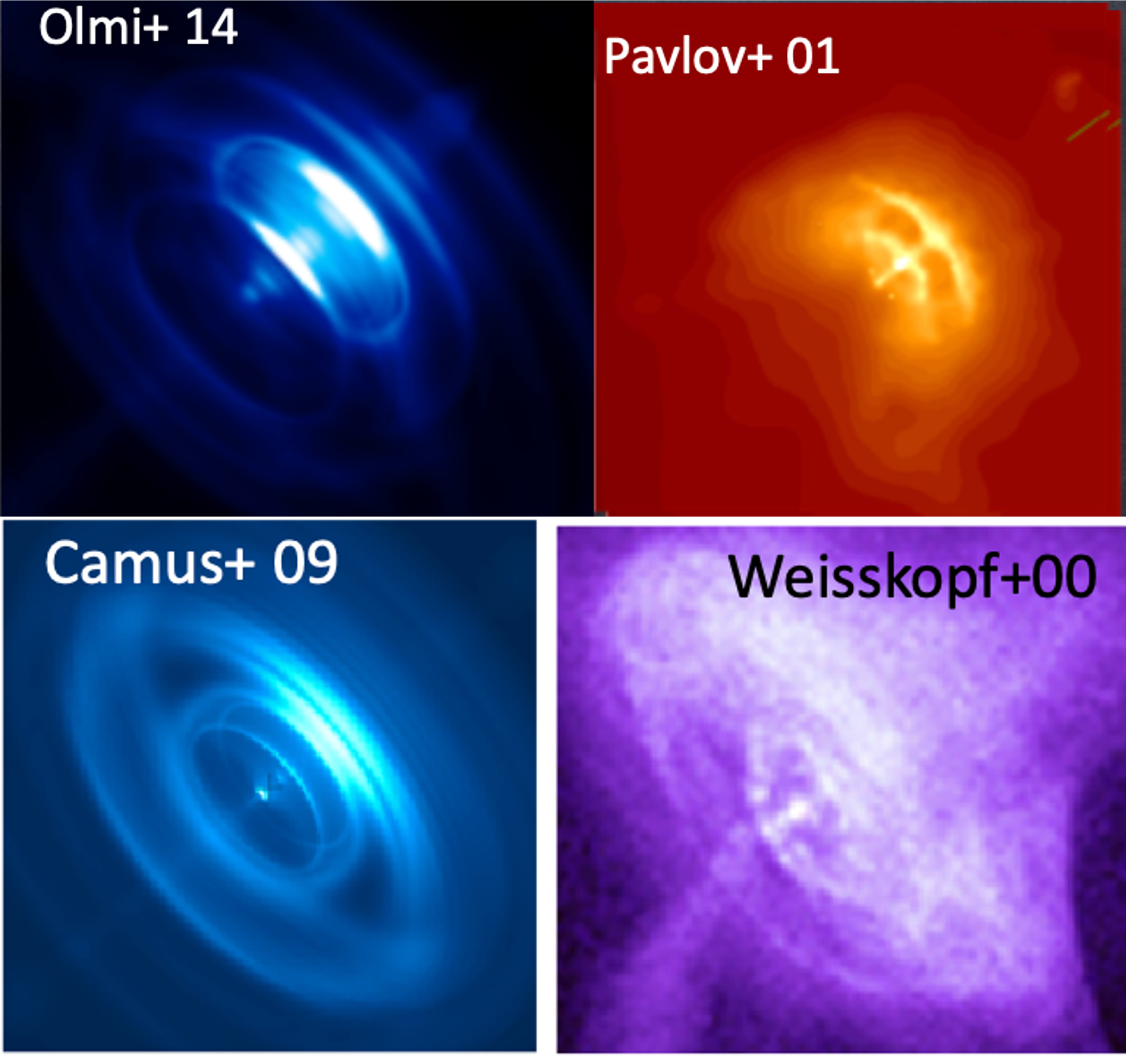}
 \caption{\footnotesize{{\bf Left panel:} the global flow velocity structure in the nebula (top) and a zoom-in on the termination shock (bottom) as derived from an axisymmetric relativistic MHD simulations with $\sigma_{TS}=0.03$ (from \cite{del-Zanna:2004}) {\bf Right panel:} X-ray synchrotron emission maps computed on top of simulation results (left, from \cite{del-Zanna:2006,Olmi:2014}) compared to actual {\it Chandra} data for Vela (top right, \cite{Pavlov:2001}) and Crab (bottom right, \cite{Weisskopf:2000}).}}
\label{fig:2DMHD}
\end{center}
 \end{figure}

These studies \cite{Komissarov:2003,Komissarov:2004,del-Zanna:2004,del-Zanna:2006, Volpi:2008,Olmi:2014} were in fact very successful at explaining the high energy of PWNe. Not only could they account for the existence of a jet originating so close from the pulsar, but they could also reproduce bright rings and {\it wisps} in the inner nebula, as an intrinsic result of the flow structure around a shock that is highly oblique at intermediate latitudes and hence leads to mildly relativistic motions in the downstream, with ensuing Doppler boosting of the emission (see Fig.\ref{fig:2DMHD}).  
A very important result of these 2D MHD studies was a revised estimate of the magnetization at the shock towards larger values. In fact, only for high enough values of $\sigma_{TS}$, the magnetic field downstream reaches equipartition and its tension is able to divert the flow towards the polar axis giving rise to a jet. The required value of $\sigma_{TS}$ depends on the nebular expansion velocity and in the case of the Crab Nebula one finds $\sigma_{TS}>0.03$ \cite{del-Zanna:2004}, an order of magnitude larger than estimates based on 1D MHD modelling \cite{KC1}.



In spite of the success at reproducing the high energy morphology of PWNe, 2D MHD modelling soon proved inadequate to fully account for the actual plasma dynamics in PWNe \cite{Volpi:2008}. Its shortcomings became immediately apparent when computing the nebular Spectral Energy Distribution. In particular, the value of $\sigma$ that best reproduces the nebular morphology fails to reproduce the multi-wavelength spectrum of the Crab Nebula: the average value of the nebular magnetic field is too low, and as a consequence, if one considers injection of a sufficiently large number of particles, so as to reproduce the high energy synchrotron spectrum, the Inverse Compton emission is largely overestimated. On the other hand, larger values of $\sigma$ lead to exceedingly large hoop stresses: too large a fraction of the flow is diverted towards the pole and the equatorial emission is too faint \cite{Olmi:2016}.
The main ingredient that the axisymmetric description is missing is the development of kink type instabilities, whose possible relevance had already been suggested in the context of explaining how to confine a wind with an initially high magnetisation at the shock \cite{Begelman98}. Indeed, when the first 3D MHD simulations of these systems were finally performed \cite{Porth:2014}, it was immediately clear that the same values of the parameters produce very different nebulae in 2D and 3D. In this latter, more realistic treatment, effective magnetic reconnection takes place downstream of the shock, reducing the hoop stresses and allowing to reproduce the nebular morphology with a larger initial value of $\sigma$. Of course in terms of reproducing both the spectrum and the morphology what is needed is a reorganisation of the field, rather than simple dissipation, so as to decrease the field tension while still keeping its strength. Long duration simulations \cite{Olmi:2016,Porth17}, in which an asymptotic self-similar solution appears to be fully reached, show that after an initial phase of continuous decrease, the average field strength reaches a constant value. However, up to now no simulation is available that can fully reproduce both the morphology and SED of the Crab Nebula and the conclusion is that situations with $\sigma_{TS}>a\ few$ must be explored with further numerical experiments.

This conclusion that the magnetization at the TS is not small has very important consequences on the particle acceleration mechanisms that can operate there. 

\subsection{Particle acceleration in PWNe}
\label{sec:partacc}
We already mentioned that a broken power-law spectrum, changing from harder to softer than $E^{-2}$ at around 500 GeV, is thought to be typical of PWNe \cite{NicJon11}. Acceleration of particles (up to PeV energies and with an efficiency close to 30\% in the Crab Nebula) is believed to occur mainly at the wind TS. This is the most relativistic shock in Nature, with a Lorentz factor estimated in the range $10^3 <\Gamma_w < 10^7$ and particle acceleration is very difficult to explain at all in this context, even with maximum energies and efficiencies much lower than the extraordinary values we infer for the Crab Nebula.
Among the various mechanisms that have been proposed, the three best studied ones are: (i) diffusive shock acceleration, or $1^{st}$ order Fermi mechanism; (ii) driven magnetic reconnection; (iii) resonant absorption of ion-cyclotron waves.

{\bf Fermi acceleration at a relativistic shock}\\
Diffusive shock acceleration is the most commonly invoked particle acceleration mechanism in Astrophysics. The basic idea is that particles gain energy every time they cross a shock front, both from upstream to downstream and vice-versa. This mechanism has been shown to guarantee efficient particle acceleration in the context of non-relativistic shock waves \cite{CaprioliSpit14}, and also at relativistic unmagnetized shock waves \cite{Spitkovsky08}, but its performance at relativistic magnetized shocks is very poor \cite{SironiKeshet15}. In fact, the essential requirement for Fermi process to work is that particles are able to cross the shock many times. This condition is very difficult to satisfy at a relativistic magnetized shock: unless the shock normal and the magnetic field direction are aligned within an angle $1/\Gamma_w$, the shock is effectively superluminal and particles cannot return from downstream, unless a very high level of turbulence is present, with $\delta B/B\gg 1$.  This requirement is easy to satisfy only if the magnetization is very low, in which case the growth of Weibel instability ensures a sufficiently high turbulence level and efficient particle acceleration. In terms of the wind magnetization, the condition for this mechanism to work is $\sigma < 10^{-3}$. In this case the shock can in principle accelerate particles, with a nearly universal spectrum $\propto E^{-2.3}$, similar to what we infer from X-ray observations of PWNe. 

However, even when this condition is satisfied, the turbulence that develops is typically small-scale turbulence. This implies that the acceleration time increases with particle energy $\propto E^2$. Therefore the particles cannot be accelerated to very high energies. This is a general problem with invoking Fermi acceleration at relativistic shocks (e.g. \cite{SironiKeshet15}). A possible way to overcome this issue would be offered by the existence of large scale turbulence of external origin in the shock vicinity, for instance MHD turbulence induced by the corrugation of the TS could be an interesting candidate in the case of PWNe \cite{Lemoine16}.\\
\\
{\bf Driven magnetic reconnection}\\
We mentioned above that the equatorial sector of the wind with alternating field lines is likely to undergo magnetic reconnection when compressed at the shock. Aside from creating a low magnetisation region where Fermi mechanism could in principle operate, reconnection can be itself accompanied by very efficient particle acceleration: if conditions are appropriate, the entire magnetic energy that is dissipated can be converted into particle acceleration, leading to very hard ($E^p$ with $1<p<2$) and extended power-law spectra. The flat spectral indices typical of the process are in the range that one generally infers from radio emission of PWNe. Moreover this mechanism is able to completely erase the Maxwellian component of the particle distribution, which on the contrary is always present in the case of Fermi acceleration and never observed in PWNe.
However, the outcome of the process in terms of spectral slope and extension depends on the initial magnetisation and pair loading of the flow. Detailed PIC simulations have shown that in the case of a particle spectrum with $p\approx 1.5$, such as observed in the Crab Nebula, in order to account for the 3 decades in particle energy that radio emission spans, one needs $\sigma \gtrsim 30$ and $\lambda/(r_L \sigma)\gtrsim$ a few tens, where $\lambda=2 \pi R_{LC}$ is the wavelength of the stripes ($R_{LC}$ is the light cylinder radius) and $r_L$ is the particle Larmor radius at the TS \cite{SironiRec11}. Assuming energy conservation along the streamlines, the latter condition can be rephrased as $\kappa \gtrsim \sigma/(1 + \sigma)(R_{TS}/R_{LC})$ where $R_{TS}$ is the TS radius. In the case of the Crab Nebula, considering the TS position coincident with the boundary of the underluminous region surrounding the pulsar at optical wavelengths, $R_{TS}\approx 0.1$ pc, which implies $\kappa>10^8$, much larger than current pulsar theories can account for \cite{TimokhinHarding19}, and also much larger than inferred from observations, as we will discuss later.\\
\\
{\bf Resonant cyclotron absorption in a ion-doped plasma}\\
An alternative mechanism that has been proposed to explain particle acceleration at relativistic transverse shock is that based on resonant absorption by electrons and positrons of the cyclotron radiation emitted by ions that are also part of the flow (e.g. \cite{HoshinoArons92}). This mechanism works for any $\sigma$, but requires that most of the energy of the pulsar wind be carried by ions. The basic physical picture is as follows: at the crossing of the TS, the sudden enhancement of the magnetic field sets the previously drifting plasma into gyration. The leptons quickly thermalize through emission and absorption of cyclotron waves, but ions with the same initial Lorentz factor (the wind is cold, so that all particles were moving with the same bulk Lorentz factor) react on time-scales that are longer by a factor $m_i/m_e$. If the wind is sufficiently cold ($\delta u/u<m_e/m_i$, with $u$ the four velocity) before the TS, the ions emit waves with large power not only at the fundamental frequency of their gyration, but up to a frequency $m_i/m_e$ times higher, which can then be resonantly absorbed by the pairs. The resulting acceleration efficiency $\epsilon_{\rm acc}$, spectral slope $\alpha$ and maximum energy $E_{\rm max}$, all depend on the fraction of energy carried by the ions $U_i/U_{\rm tot}$. PIC simulations show a wide variety of values: $\epsilon_{\rm acc}=$ few (30)\%, $p>3 (<2)$, $E_{\rm max}/(m_i \Gamma_w c^2)=0.2 (0.8)$ for $U_i/U_{\rm tot}=0.6 (0.8)$ \cite{AmatoArons06}. Once again the pulsar multiplicity and the related question of the origin of radio particles play a crucial role.

\subsection{Constraints on particle acceleration mechanisms based on the nebular dynamics}
\label{sec:partaccconstr}
One firm conclusion that could be drawn from numerical studies is that the wind magnetisation of the shock must be much larger than previously thought: in the case of the Crab Nebula one may estimate $\sigma_{TS}> a\ few$. This has profound implications on the viable acceleration processes at the shock: as already discussed, the Fermi process cannot operate in regions where the plasma has $\sigma>0.001$. Of course at the TS the local magnetisation of the flow will depend on latitude above the equator and substantially lower values can be found in the equatorial sector that hosts the current sheet. However when we evaluate the fraction of the flow that satisfies the condition $\sigma<0.001$, we find it to be insufficient to account for the Crab Nebula X-ray emission \cite{Amato14}. This conclusion might be changed if effective focusing of particle trajectories towards the equator occurs and the level of magnetic turbulence downstream of the shock is high enough \cite{Giacinti18}.

In any case, the possibility that different acceleration mechanisms operate at different latitudes along the TS is an interesting one to consider. This kind of investigation is primarily prompted by observations, aside from any theoretical consideration. In fact, \cite{Schweizer13,Bietenholz:2004} performed an investigation of the Crab Nebula {\it wisps} at multi-wavelengths. The result of such analysis is that the {\it wisps} are not coincident at radio, optical and X-ray frequencies. Within a MHD description of the flow, the wisps are interpreted as regions of enhanced magnetic field and fast flow motion towards the observer (increasing the Doppler boosting effect). Therefore, differences in their appearance and behaviour can only be accounted for if the particles responsible for the emission at different frequencies are injected at different locations along the shock front. In particular, detailed analysis of synthetic emission maps, computed on top of the MHD flow with different prescriptions on the injection of particles in different energy ranges, showed that the observed wisps behaviour is best reproduced if optical/X-ray emitting particles are injected mostly in the vicinity of the equator and radio emitting particles are injected either at higher latitudes or everywhere along the shock front \cite{Olmi:2015}. 
In fact the observed behaviour of the wisps at radio wavelengths is even compatible with a more extreme hypothesis, namely that the particles responsible for radio emission are uniformly distributed in the nebula rather than accelerated at the shock front and then advected with the flow. Available radio images of the Crab Nebula do not allow to discriminate between a scenario in which radio emitting particles are accelerated at the TS alone and one in which they are uniformly distributed throughout the nebula \cite{Olmi:2014}. The only scenario that radio data allow us to exclude is one in which the particles responsible for the emission are fully relic, namely injected in the nebula at early times (as suggested e.g. by \cite{Atoyan99}) and then advected with the flow. A relic origin of the low energy particle population is instead compatible with observations if one assumes that they are not simply advected, but rather effectively scattered and reaccelerated by turbulence (Fermi II type acceleration) or magnetic reconnection.

This latter result has profound implications on the estimate of the pulsar multiplicity, $\kappa$. Indeed, for typical particle spectra in PWNe, radio emitting particles are dominant by number (see Fig.~\ref{fig:sketch}). If they are part of the pulsar wind, they determine the value of $\dot N$ that must be supplied by the wind, and hence the values of $\Gamma_w$ and $\kappa$ in Eq.~\ref{eq:winden}. In the case of the Crab Nebula one finds $\dot N\approx 10^{40} {\rm s}^{-1}$ and hence $\kappa\approx 10^6$ and $\Gamma_w\approx 10^4$. On the other hand, if these particles are not currently part of the pulsar wind, but were rather injected in the nebula at some earlier time (when the pulsar was much younger and more energetic \cite{Atoyan99}) or even extracted from the thermal plasma in the supernova ejecta, then the flow parameters would be determined by the population of X-ray emitting particles. These require $\dot N\approx 10^{38.5} {\rm s}^{-1}$ and hence $\kappa\approx 10^4$ and $\Gamma_w \approx few \times 10^6$. 
These estimates offer different constraints on the viable mechanisms of particle acceleration at the TS. If radio emitting particles are part of the pulsar wind, the estimated value of $\kappa$ is larger than current models of pair creation in the magnetosphere can account for \cite{TimokhinHarding19}, yet below the value ($10^8$) that would allow acceleration by driven magnetic reconnection to produce a particle spectrum such as that observed in Crab \cite{SironiRec11}. However, it should be kept in mind that $\kappa>10^8$ was derived based on particle density at the TS position in the equatorial plane. We now know that the TS is much closer to the star along the polar axis, and since the density of the wind scales $\propto r^{-2}$ (with $r$ the distance from the pulsar), a value $\kappa\lesssim 10^6$ would be sufficient if reconnection was taking place there. The problem is that we expect high latitude stripes only in the case of an orthogonal rotator.
If $\kappa\approx 10^6$, the pairs are so many that, even if present in the wind at the level of a Goldreich \& Julian flux, ions could not dominate its energy content and resonant cyclotron absorption of ion waves could not work as an acceleration mechanism for pairs. On the other hand, if radio emitting particles are not being supplied by the pulsar, $\kappa$ is within the range of values that magnetospheric
models predict and is also such that ions can be energetically dominant and accelerate pairs at the shock. It is interesting to notice that the presence of ions in the pulsar wind would also ease the conversion of Poynting flux to kinetic energy flux that must occur in the wind between the light cylinder and the TS \cite{KirkGiacinti19}. In addition, ion dominance is the requirement for pulsar winds to be the primary contributors of UHECRs (see Sec.\ref{sec:uhecrs}).


\subsection{News from UHE gamma-rays}
\label{sec:PeV}
The broad implications that the presence of ions in pulsar winds would have make it mandatory to look for their signatures in all available channels. Aside from high energy neutrino detection \cite{Amato:2003}, that would obviously provide smoking-gun evidence, the other viable channel for PWN protons to show up is through production of multi-TeV gamma-rays deriving from the decay of neutral pions produced in nuclear collisions with the gas in the SN ejecta. 
\begin{figure}[h!!!!]
 \begin{center}
 \includegraphics[width=.9\textwidth]{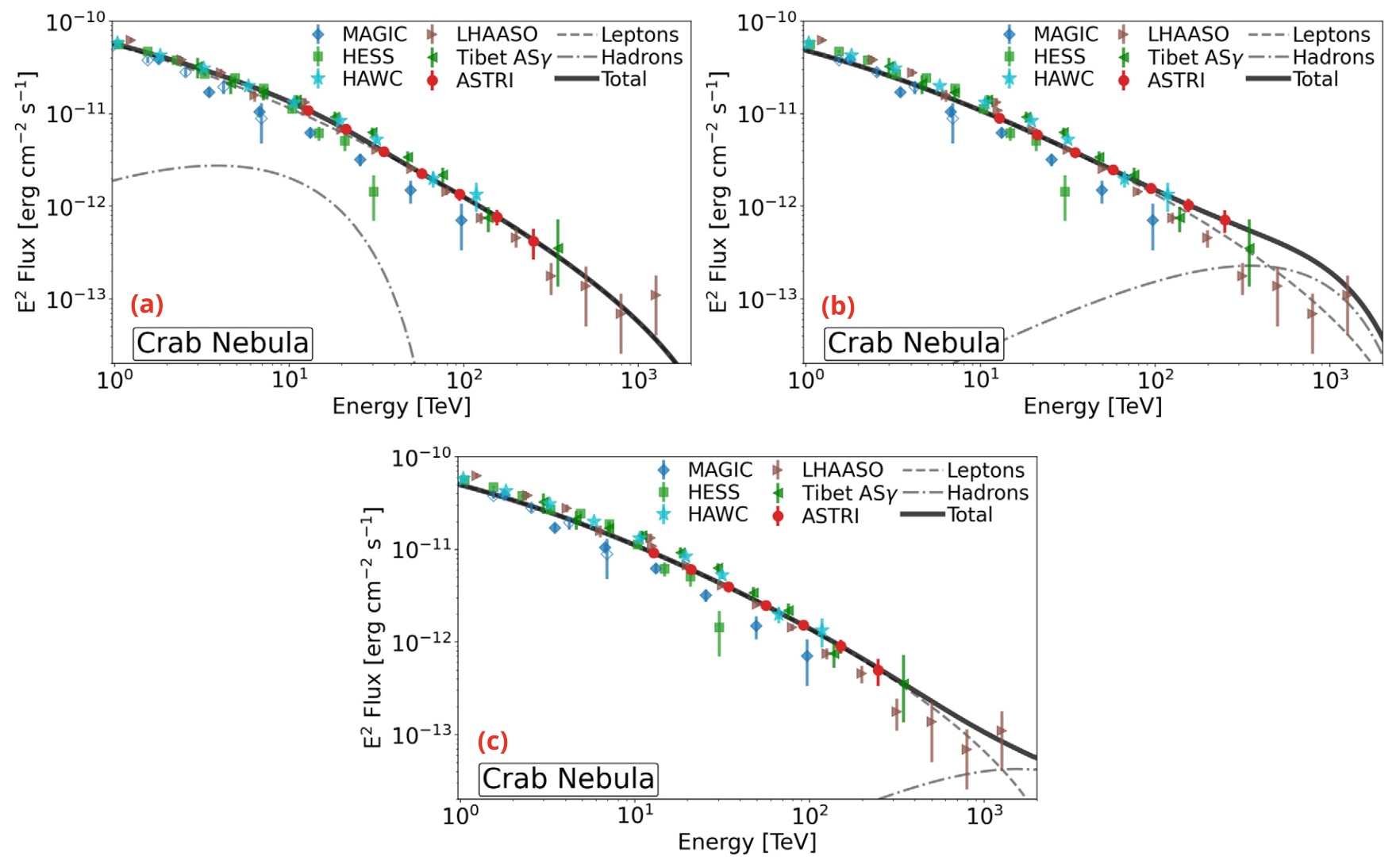} 
 \caption{\footnotesize{Spectrum of the Crab Nebula modeled with a hadronic component characterized by: (a) $\Gamma_w=10^5$ and 15\% of the wind energy in protons; (b) $\Gamma_w=5 \times 10^6$ and 4\% of the wind energy in protons; (c) $\Gamma_w=2 \times 10^7$ and 4\% of the wind energy in protons. Grey lines indicate the leptonic (dashed line) and hadronic (dot-dashed line) components, black line shows the total emission. Data from different instruments are shown with various symbols/colors as specified in the different panels. Figure taken from \cite{ASTRISci}.}}
\label{fig:astri}
\end{center}
 \end{figure}
In the case of the Crab Nebula, which is so bright in ICS up to very high energies, this spectral contribution is only expected to become detectable above $100- 150$ TeV, where IC scattering emission starts to be suppressed by the Klein-Nishina effect.

Very recently LHAASO obtained the record breaking detection of $>$PeV photons from this source \cite{cao2021peta}, opening up a window to finally see the possible emergence of the hadronic contribution. In fact, the increasing uncertainties above 500 TeV make LHAASO spectrum still consistent with a purely leptonic origin of the emission. On the other hand, taken at face value, the LHAASO data seem to suggest that a new component might be showing up at the highest energies. This new component is consistent with a quasi-monochromatic distribution of protons with energy around 10 PeV (as discussed in \cite{ASTRISci}) and shown in Fig.\ref{fig:astri}. This is exactly what would be expected by models assuming that protons are part of the wind emanating from the Crab pulsar with a Lorentz factor $\Gamma_w\approx 10^7$: in this case their Larmor radius in a $100\,\mu$ G is of order $R_{\rm TS}$, so large that their energy distribution would not be much altered at the shock \cite{AmatoArons06}. Decisive insight will hopefully be provided by next generation IACTs with good sensitivity beyond 100 TeV as the CTA SSTs (Small Size Telescopes) in the southern hemisphere and ASTRI Mini-Array in the north. 
 
The Crab Nebula is not the only source and likely not the only PWN to have been detected at UHE. LHAASO \cite{LHAASO_EHE} has also detected about ten more UHE emitters in the Galaxy (partially overlapping with the sources already detected by HAWC \cite{HAWC100} beyond 56 TeV). For the majority of these sources, the distance between the center of the emission and the nearest pulsar is less than or comparable with the instrument PSF, so that there have been suggestions that these sources are all associated with pulsars and leptonic in nature \cite{Breuhaus2021}. 
Since at PeV energies the only available target for ICS emission is the CMB, which requires an electron of energy $E_e\approx 2.15\ E_{\gamma, PeV}^{0.77}$ PeV, to emit a photon of energy $E_\gamma$ ($E_{\gamma,PeV}$ is the photon energy in units of PeV, \cite{LHAASO_EHE}). On the other hand, if the radiation were instead of hadronic origin, an even higher particle energy would be required, being the energy of photons produced via p-p collisions about 1/10 of the parent proton energy.

In fact there is a very simple and solid argument that can be used to check if LHAASO sources can indeed be associated with pulsars: the pulsar potential drop should be high enough to allow the acceleration of particles to the needed energy. This statement might seem surprising at first sight, after discussing at length particle acceleration at the TS: we mentioned the potential drop as a limit for acceleration in the magnetosphere, not at the shock. However, one can easily prove that $\Delta \Phi_{pc}$ in Eq.\ref{eq:dphingj} is an absolute limit to the energy of particles accelerated anywhere in the pulsar-PWN system. The argument is as follows.
The maximum achievable energy at the TS is the potential drop:
\be
E_{\rm max,TS}=Z e \xi_E B_{TS} R_{TS}
\label{eq:emaxgen}
\ee
where $Z\ e$ is the particle charge and $\xi_E\leq 1$ is the ratio between electric and magnetic field in the acceleration region, typically $\xi_E\approx v/c$ with $v$ the local flow velocity. Pulsar winds, being relativistic outflows, can in fact realize $\xi_E\sim 1$\footnote{in fact, $\xi_E>1$ is inferred occasionally in Crab, during the so-called gamma-ray flares. We will not discuss these here, but an excellent review is offered by \cite{BlandBuehler}}. 
On the other hand, the magnetic field at the TS, $B_{TS}$, is bound to be a fraction $\xi_B< 1$ of the overall pressure at the TS, which is in turn related to the pulsar wind energy flux:
\be
\frac{B_{TS}^2}{2 \pi}=\xi_B \frac{\dot E}{4 \pi R_{TS}^2 c}\ .
\label{eq:BTS}
\ee

\begin{figure}[h!!!!]
 \begin{center}
 \includegraphics[width=.8\textwidth]{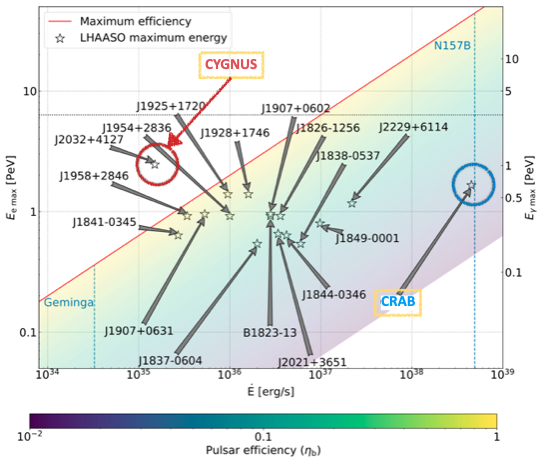} 
 \caption{\footnotesize{Maximum electron energy derived from the LHAASO spectra \cite{LHAASO_EHE} vs. spin-down power of the co-located pulsars. The right y-axis shows the corresponding gamma-ray energy. The colored area shows the values for $\xi_E \xi_B^{1/2}$ ranging from 0.01 to 1, with the red line indicating the limiting value corresponding to maximally efficient acceleration $\xi_E=\xi_B=1$. The dotted black line marks the upper limit to the maximum energy for young pulsars with a large magnetic field of 100 $\mu$G. The blue dashed horizontal lines show the predicted values for PWNe associated with Geminga and N157B. The source in Crab is highlighted as the one reaching energies corresponding to the lowest fraction of the full potential drop available (see text). The pulsar in Cygnus, J2032+4127, is highlighted as the one whose potential drop is too low to account for the observed PeV emission.
}}
\label{fig:psrpev}
\end{center}
 \end{figure}
Combining these two equations results in:
\be
E_{max,abs}\approx 2 Z e\ \xi_E \xi_B^{1/2} \sqrt{\dot E/c}\approx 2 PeV\ Z\ \xi_E \xi_B^{1/2} \dot E_{36}^{1/2}\ .
\label{eq:TSdrop}
\ee
 
Eq.\ref{eq:TSdrop} immediately shows that only pulsars with $\dot E\gtrsim 10^{36}$ can accelerate particles to high enough energies to produce PeV radiation. This argument, that only requires knowledge of the pulsar $\dot E$, was used by \cite{deona} to check whether LHAASO sources could all be associated with pulsars observed in the vicinity of the UHE gamma-ray hotspots. The results are shown in Fig.\ref{fig:psrpev}, adapted from \cite{deona}.
 
Fig.\ref{fig:psrpev} shows that indeed most of the LHAASO sources can in principle be associated with pulsars, with the noticeable exception of the emission detected from Cygnus. In this case the PeV emission is likely to come from a different type of source, possibly the Young Massive Star Cluster also present in the region \cite{FelixWinds}. Another noticeable detail in the figure is the fact that the Crab Nebula, often taken as the paradigm of an efficient accelerator, seems to achieve in fact a small fraction of the full potential available. The reason for this is that in this source the maximum electron energy is not limited by $E_{max,abs}$ in Eq.\ref{eq:TSdrop} but rather by radiation losses. Indeed, another, occasionally more stringent, limitation to the maximum achievable energy for an accelerator comes from the condition 
\be
t_{acc}=\frac{E}{e \xi_EBc}\leq t_{loss}=\frac{6 \pi (mc^2)^2}{\sigma_T c B^2 E}\ ,
\label{eq:losslim}
\ee
which defines
\be
E_{max,loss}\approx 6\ PeV\ \xi_E^{1/2} B_{-4}^{1/2}\ .
\label{eq:emaxloss}
\ee
For young and bright sources, with local fields $\gtrsim 10^{-4}$ G, as is the case for the Crab Nebula TS, this is a more stringent limit on the maximum energy, which accounts for its location in the plot in Fig.\ref{fig:psrpev}. In addition, Eq.\ref{eq:emaxloss} highlights two things: 1) among galactic sources, only pulsars are likely to host PeV leptons, while other powerful accelerators, with non-relativistic outflows, are bound to have $\xi_E\ll1$ (e.g. typically $\xi_E\approx 10^{-2}$ for SNRs) and cannot bring leptons to such high energies; 2) in the case of pulsars, the detection of UHE radiation and its interpretation as leptonic immediately puts constraints on the magnetic field strength at the termination shock \cite{deona}, and hence can be used to constrain the pulsar wind properties.

For evolved systems, with magnetic field strengths in the $10s$ of $\mu$G range and $\dot E>10^{35}$ erg/s, Eq.\ref{eq:emaxloss} gives an energy larger than the polar cap potential drop, and the latter is the actual constraint. These systems are the most abundant galactic TeV emitters and likely the main contributors of CR positrons above $\sim 20$ GeV. The last section is devoted to a short summary of their properties and of the hottest related open questions.


\subsection{Evolved systems and particle escape}
\label{sec:escape}
Aside from being guaranteed leptonic PeVatrons and potential sources of protons at the highest galactic energies and possibly even at UHE, PWNe had previously called the attention of the CR physics community as possible primary contributors of cosmic ray positrons. In the last decade, it has become evident, with ever increasing statistical significance, that the ratio between CR $e^+$ and $e^-$ increases with energy above $\sim 10$ GeV \cite{PAMELApos,AMS02pos}. This finding is in contrast with the expectations of models considering positrons as pure secondary products of cosmic-ray interactions during propagation through the Galaxy and has prompted many efforts to provide an explanation. A pulsar related origin of the unexpected positrons is among the most popular suggestions. In particular, evolved PWNe seem a most promising candidate \cite{Evoli21}.

\begin{figure}[h!!!]
\begin{center}
\includegraphics[width=0.41\textwidth]{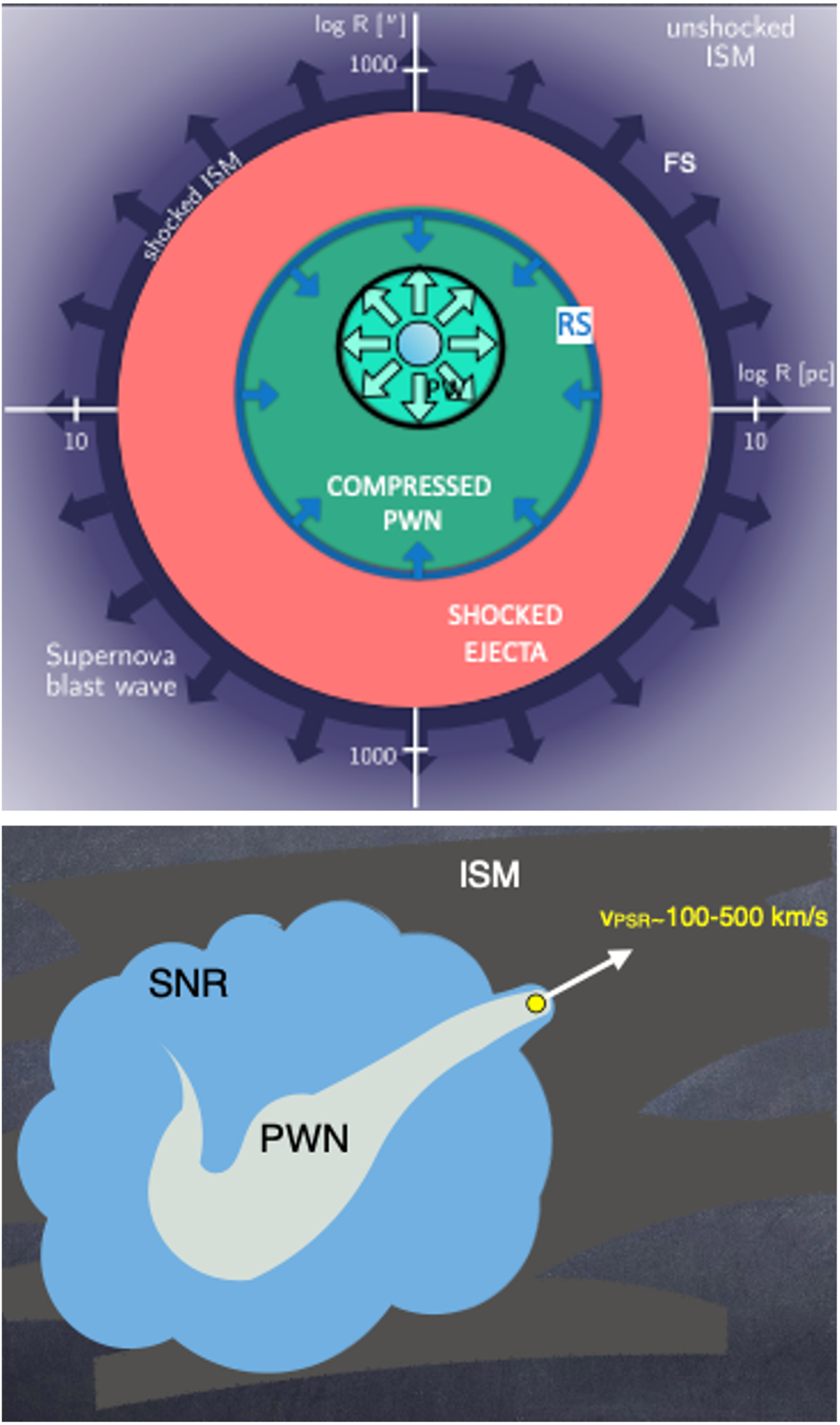}  
\hspace{1cm}
\includegraphics[width=0.49\textwidth]{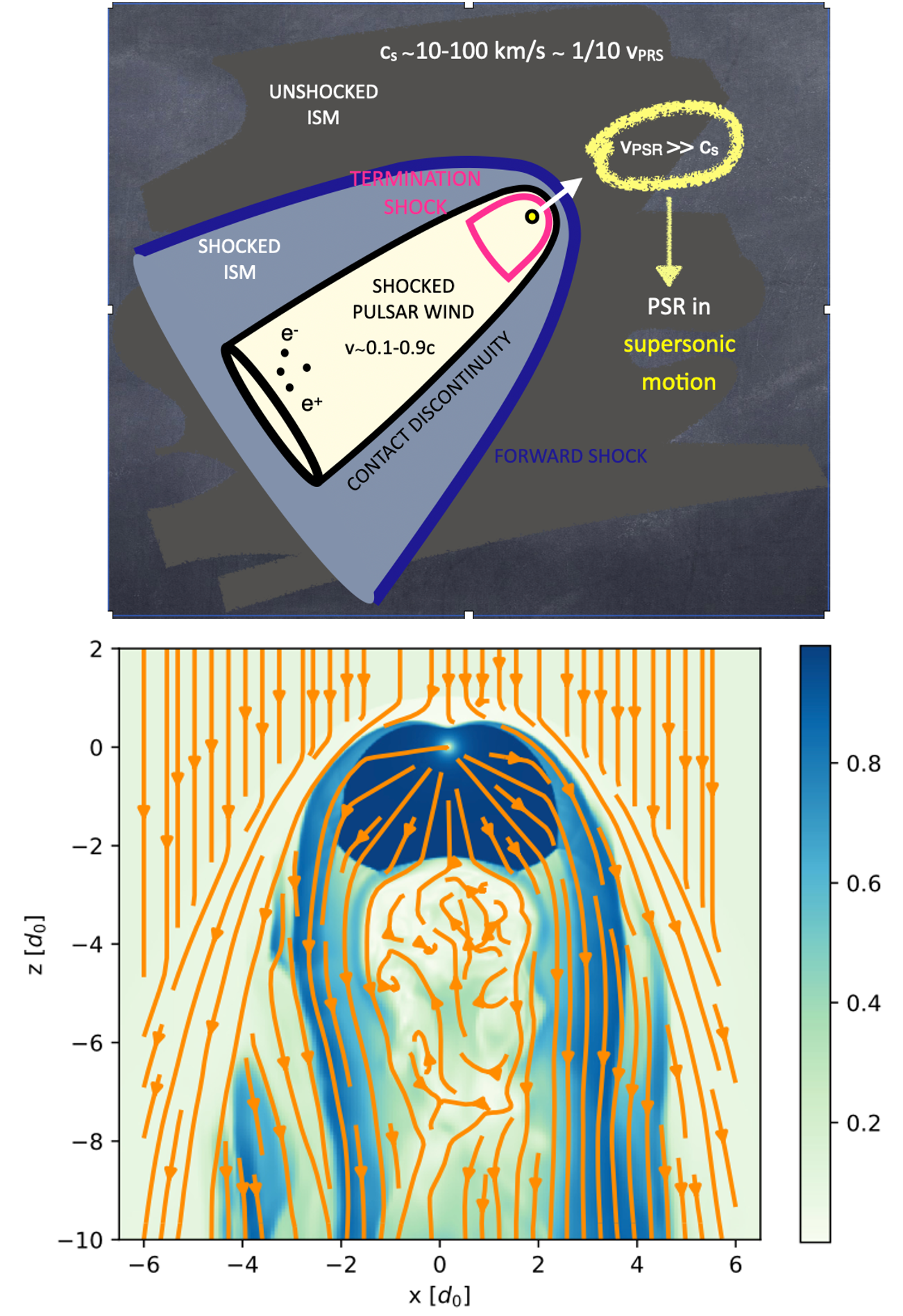}
\caption{\footnotesize{Typical evolution of a PWN. The top panel shows the pulsar moving away from the SN explosion center. After few tens of kyr the pulsar leaves the parents SNR (bottom left panel) and creates a bow shock nebula (top right panel). The bottom right panel (from \cite{OlmiBucc1}) shows the streamlines of the flow in direction and magnitude (the color scale) from a 3D simulation of a bow shock. Most of the plasma flows away from the pulsar towards the back of the bow shock.}}
\label{fig:bowsketch}
\end{center}
\end{figure}

The positron excess is detected at the energies that characterise the radio emitting particles in these systems and the excess is best explained with a spectrum $\sim E^{-1.5}$ or in any case harder than $E^{-2}$, which is what typically inferred from radio emission of PWNe. These particles, however, have small Larmor radii compared to the size of the system and cannot easily escape as long as the PWN is young and inside its parent SNR. However the high average proper motion of the pulsar population causes the nebula to leave the parent SNR within a time-scale of order few $\times 10^4$ yr, and at that point the nebula is likely to become a Bow Shock PWN: the pulsar supersonic motion drives a shock in the ISM, confinement of the pulsar wind is guaranteed by the ram pressure of the ISM in the front, but the plasma is free to leave the system from the  back. A sketch of the system evolution with time is shown in Fig.\ref{fig:bowsketch}. In particular the flow streamlines from a 3D relativistic MHD simulation \cite{OlmiBucc1} are shown in the bottom right panel. 

At that stage, particles are not only free to leave the PWN from the tail of the bow shock, but also likely to escape from close to the bow shock head, thanks to the development of instabilities that break the contact discontinuity and let the relativistic plasma flow freely in the ISM. Proof of such a phenomenon is directly observed at least in a few systems showing extended trails of X-ray emission developing perpendicular to the bow shock and inferred to be due to particles with energies extremely close to the polar cap potential drop of the parent pulsar propagating in a field that is about a factor of 10 larger than the average field in the ISM \cite{HuiBecker07,Pavan16}.

On the other hand, recent observations have cast some doubts on the effective escape of leptons from the pulsar surroundings. Indeed the detection by HAWC \cite{HAWCHalo} of extended halos of multi-TeV emission around Geminga and PSR B0656+14 leads to infer that the diffusion coefficient in the vicinity of these systems is much reduced with respect to the galactic average, and particles might undergo relevant energy losses before leaving the region and becoming part of the galactic CR pool. 

The physics governing particle escape from evolved PWNe is a subject that is only just starting to be investigated and several aspects are mysterious and fascinating. Some open questions are the following.

First of all both the formation of halos and of X-ray trails suggests that the escaping particles are somehow altering the medium in which they propagate, amplifying the turbulence level in the case of halos and even increasing the overall magnetic field strength in the case of trails. Recent numerical studies undertook the problem of tracing particle trajectories on top of the MHD flow structure provided by numerical simulations \cite{Olmi&Bucciantini:2019_3}. These studies, while preliminary, have highlighted several interesting properties of the escape: 1) a fraction of particles can leave the system not only from the back, but actually also from the head, forming outflows in a direction that depends on the magnetic field structure in the nebula and in the ISM; 2) particles at the highest energies can in principle leave the system with high efficiency and almost isotropically, while the fraction of escaping particles and the angular extent of the flow progressively decrease with decreasing particle energy; 3) electrons and positrons flow away along different trajectories, so that the flow is charge-separated at some level and carries a non-null electric current. This latter aspect suggests that the particles might induce current driven instabilities in the ISM and these might be at the origin of the amplified magnetic field deduced from observations.

Building on these preliminary results, the different appearance of the two phenomena, halos and X-ray trails, needs to be understood in terms of basic properties of the parent system and/or of the local ISM. Such a study is especially important for cosmic ray physics in two respects: 1) to determine the particle spectrum eventually released by Bow Shock PWNe in the ISM and if these sources can actually account for the positron excess or leave room for additional phenomena; 2) to determine how wide spread in the Galaxy the regions of reduced diffusivity associated with pulsar haloes are, since their abundance could in principle affect the global CR transport in the Galaxy.

\section{Conclusions}
\label{sec:concl}
In the last 20 years the connection of pulsars and PWNe with CR physics has become increasingly tight and it is now clear that understanding these sources is essential also to understand several aspects of CR origin and propagation. The open questions are many, as I have tried to highlight, and exciting times are ahead in the quest for answers.

\acknowledgments
The author acknowledges INAF that supported this work through grant PRIN-INAF 2019.


\bibliography{VarennaBiblio}

\end{document}